\documentclass[aps,groupedaddress,twocolumn,pdftex]{revtex4}
\usepackage{amsfonts,amsmath,graphicx}
\usepackage[letterpaper,centering,top=1.5in,bottom=1.0in,left=.5in,right=.5in]{geometry}
\newcommand{\be}{\begin{equation}}
\newcommand{\ee}{\end{equation}}
\newcommand{\ba}{\begin{eqnarray}}
\newcommand{\ea}{\end{eqnarray}}

\def\bea{\begin{eqnarray}}
\def\eea{\end{eqnarray}}
\def\ben{\begin{eqnarray*}}
\def\een{\end{eqnarray*}}

\def\>{\rangle}
\def\<{\langle}

\begin{document}

\title{A Two-Dimensional Lattice Ion Trap for Quantum Simulation}
\author{Robert J. Clark \footnote[1]{Both authors contributed equally to this work.} \footnote[2]{Email: robclark@mit.edu}, Tongyan Lin \footnotemark[1] \footnote[3]{Email: tongyan@physics.harvard.edu}, Kenneth R. Brown, and Isaac L. Chuang}

\affiliation{Research Laboratory of
Electronics \& Department of Physics\\ Massachusetts Institute of
Technology, Cambridge, Massachusetts 02139, USA}

\date{\today}

\begin{abstract}
\noindent
Quantum simulations of spin systems could enable the solution of problems which otherwise require infeasible classical resources. Such a simulation may be implemented using a well-controlled system of effective spins, such as a two-dimensional lattice of locally interacting ions. We propose here a layered planar rf trap design that can be used to create arbitrary two-dimensional lattices of ions. The design also leads naturally to ease of microfabrication. As a first experimental demonstration, we confine $^{88}$Sr$^+$ ions in a mm-scale lattice trap and verify numerical models of the trap by measuring the motional frequencies. We also confine 440~nm diameter charged microspheres and observe ion-ion repulsion between ions in neighboring lattice sites. Our design, when scaled to smaller ion-ion distances, is appropriate for quantum simulation schemes, e.g. that of Porras and Cirac (PRL \textbf{92} 207901 (2004)). We note, however, that in practical realizations of the trap, an increase in the secular frequency with decreasing ion spacing may make a coupling rate that is large relative to the decoherence rate in such a trap difficult to achieve. 

\end{abstract}
\pacs{}
\maketitle


\section{Introduction}

Elucidating the low temperature properties of magnetic materials remains a challenge to  computational physics. The combinatorial number of states in a many-spin system, as well as the quantum-mechanical nature of the materials, makes simulations difficult. These effects are strongly dependent on the geometry of the spin system and the existence of defects in the spin lattice \cite{Sachdev:book, Diep:book, Moessner:01}.  Current tractable models of these systems rely on insights into the geometry that yield models with fewer degrees of freedom \cite{Jiang:05}.

An alternative to classical numerical computation is quantum simulation, in which one well-controlled quantum system is used to simulate the properties of another. The type of simulation possible is currently limited by the controllability of the system and the number of degrees of freedom; a tradeoff between the two has existed in every experimental implementation of quantum simulation to date. Small molecule NMR \cite{Cory:99,Laflamme:04,Suter:04,Brown:06} and experiments with a small number of trapped ions \cite{Wineland:98b,Leibfried:02,Schaetz:08} have demonstrated the principle of universal quantum simulation for small systems. On the other hand, atoms in optical lattices have been used to simulate large systems but with limited control \cite{Cataliotti:01,Paredes:04,Weiss:04,Greiner:02}.

Theoretical proposals for quantum simulation of spin systems have included the addition of controls to current optical lattice experiments \cite{Hofstetter:02, Duan:03, Sorensen:05} or the use of many trapped ions \cite{Jane:02, Milburn:05, Porras:04,Porras:04BEC, Deng:05, Porras:06}. Porras and Cirac show in Ref. \cite{Porras:04} a way to generate an effective antiferromagnetic 2D Ising interaction in an array of ions. Such a simulation has recently been performed for the ferromagnetic case with two ions in a linear trap \cite{Schaetz:08}. However, the observation of spin frustration requires a 2-D array of ions. One of the principal technical questions of quantum simulation research has been how to realize such a two-dimensional lattice of ions. A 2-D array of ions has been realized in a Penning trap by NIST \cite{Itano:98}, but inconveniently the crystal rotates due to the crossed \textbf{E} and \textbf{B} fields. Arrays of electrons in individual Penning traps \cite{Stahl:05} have also been proposed as one solution, although the schemes of quantum simulation mentioned above for trapped ions are not directly applicable. A recent proposal \cite{Chiaverini:08} proposes a scheme for doing quantum simulations in 2-D arrays of microtraps using localized electromagnetic fields and magnetic field gradients. This scheme does not use pushing forces from lasers, as in Ref.~\cite{Porras:04}, eliminating errors due to sponataneous scattering and simplifying the optics needed considerably. 

\begin{figure}[t]
\begin{center}
\includegraphics[width=.45\textwidth]{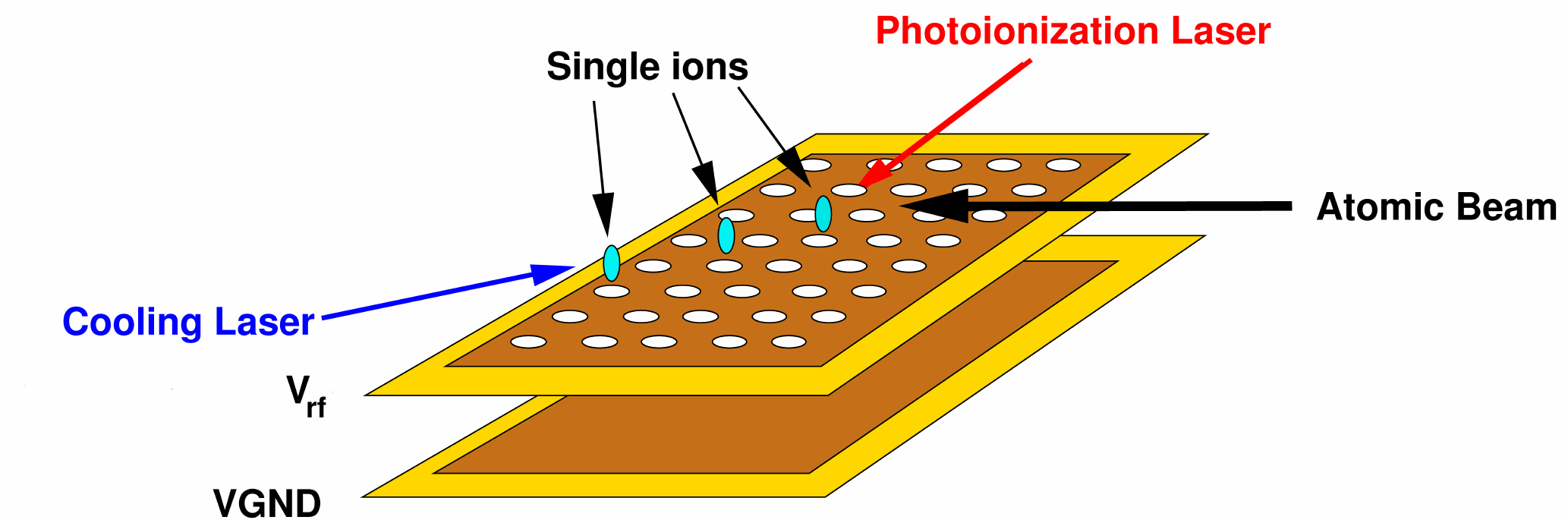}
\caption{Schematic of the lattice trap. An array of traps is produced by a single rf electrode with a regular array of holes, 
mounted above a grounded electrode. Ions will preferentially
be loaded from a broad atomic beam at the intersection of
the cooling and photoionization lasers.}
\label{fig:atomictrap}
\end{center}
\end{figure}


 In this paper, we present a layered planar rf electrode geometry that creates a 2-D ion lattice for quantum simulation of Hamiltonians in 2-D. The ion trap consists of a planar electrode with a regular array of holes, held at a radiofrequency (rf) potential, mounted above a grounded planar electrode. A single ion is trapped above each hole in the rf electrode (Fig.~\ref{fig:atomictrap}). Ions will be preferentially loaded above the trap electrode at the intersection of the Doppler cooling and photoionization beams, allowing the user to write an arbitrary 2-D lattice structure. Also, in this Paul trap-based scheme, we avoid the large Zeeman shifts associated with Penning traps, as well as the rotation of the ion crystal. Understanding the basic properties of this trap is important not only for Porras and Cirac's proposal, but also for related methods such as Chiaverini and Lybarger's magnetic field gradient approach, and any ion trap quantum simulation protocol that relies on a fixed 2-D array of trapped ions. Our research here focuses on three trap properties needed for quantum simulation: the  ability to stably confine ions, predictable trapping potentials, and measurable interactions of ions located in adjacent wells. Therefore, in this work, we ask the following questions: 1) How well can our design be used to trap an array of ions? 2) How well do numerical models of the trap match its observed properties? and 3) What ion-ion interactions between ions in neighboring wells can be observed? 

The paper is organized as follows. We first present a theoretical model of the lattice trap (Section II), and then report on two demonstrations of the trap. In the first, we confine $^{88}$Sr$^+$ ions and test numerical models of the trap by measuring the motional frequencies of the ions (Section III). We then use $440$~nm-sized charged microspheres to measure ion-ion repulsion (Section IV). In Section V, we estimate the ion-ion spacing required in our trap to realize a quantum simulation. In Section VI, we summarize our results and discuss future work. 


\section{Lattice Trap Design \label{Sec:Paul}}

\begin{figure}[b]
\begin{center}
\includegraphics[width=.25\textwidth]{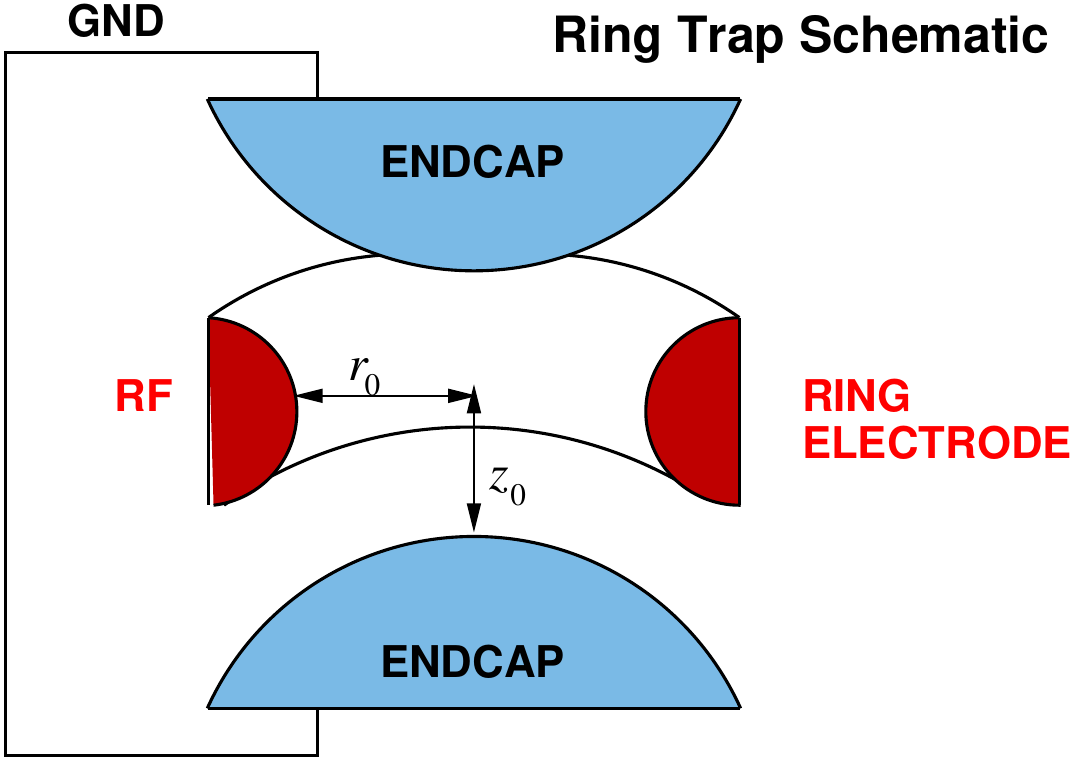}
\caption{Cross section of a ring trap. Ions are confined at the center of the trap. The ring is held at an alternating RF potential relative to the endcaps. Here $r_0 = \sqrt{2}  z_0$ and the endcaps and ring electrode are hyperbolically shaped.}
\label{fig:ringtrap}
\end{center}
\end{figure}

Our lattice trap is an extension of the three-dimensional ring Paul  trap~\cite{Ghosh:book}. Following this reference, we first review the theory of the ring trap. The ring electrode geometry is shown in Fig.~\ref{fig:ringtrap}. An alternating voltage of the form $V_{rf} = V \cos \Omega t $ is applied to the ring electrode and the endcaps are grounded. 
The equations of motion for an ion in the ring trap are a set of Mathieu equations which have regions of stability depending on the dimensionless parameters $a = \frac{8QU_0}{m r_0^2 \Omega^2}$ and $ q  = \frac{2 Q V}{m r_0^2 \Omega^2}$. $Q$ and $m$ are charge and mass of the ion, $U_0$ is any DC voltage applied to both endcaps of the ring trap, and the distance $r_0$ is the distance from the trap center to the rf electrode (as shown in Fig.~\ref{fig:ringtrap}). When $U_0 = 0$ and the system is in vacuum, the condition for stability is $q < .908$. 

We assume the trajectory of a trapped ion is well approximated by a slow secular motion superposed with a rapid oscillation, the micromotion, due to the oscillation in the potential $V_{rf}$. For $U_0= 0$, time-averaging the ion motion in the secular approximation ($q \ll 1$) gives the following quasi-static pseudopotential which governs the secular motion:
\begin{equation}	
	\Psi (\vec x) = \frac{Q^2}{4m \Omega^2} |\vec \nabla \Phi (\vec x)|^2.
\label{eq:pseudopotential}
\end{equation} Here $\Phi(\vec x)$ is the electrostatic potential when the drive voltage $V$ is applied to the ring electrode. 

The lattice trap can be thought of as a planar array of ring traps. Ions are confined in a 2-D lattice of potential wells. As discussed in Sec. I, this trap comprises two layers: an rf plate (extended ring electrode), and a grounded plane beneath it. At the center of each  trap, the electric field associated with the electrostatic potential $\Phi (\vec x)$ is 0. Assuming approximate rotational symmetry in the plane of the trap, $\Phi (\vec x)$ has a multipole expansion:
\begin{equation}
	\Phi (\vec x) = V\frac{r^2 - 2 z^2}{r_1^2} + \alpha V \frac{2z^3 - 3zr^2}{r_1^3}
\label{eq:harmonicpot}
\end{equation} where $r$ is the radial distance from the central axis of the lattice site and $z$ is the distance along the central axis. The above expression is valid for infinite lattices, but for lattice traps containing many ions the potential will be correct near the center lattice site. The $z = 0$ plane is defined such that it coincides with the point of null electric field. Eq.~\ref{eq:harmonicpot} defines two constants which depend on the trap geometry: $r_1$, with dimension of length, and $\alpha$, which is dimensionless. 

\begin{figure}[b]
\begin{center}
\includegraphics[width=.45\textwidth]{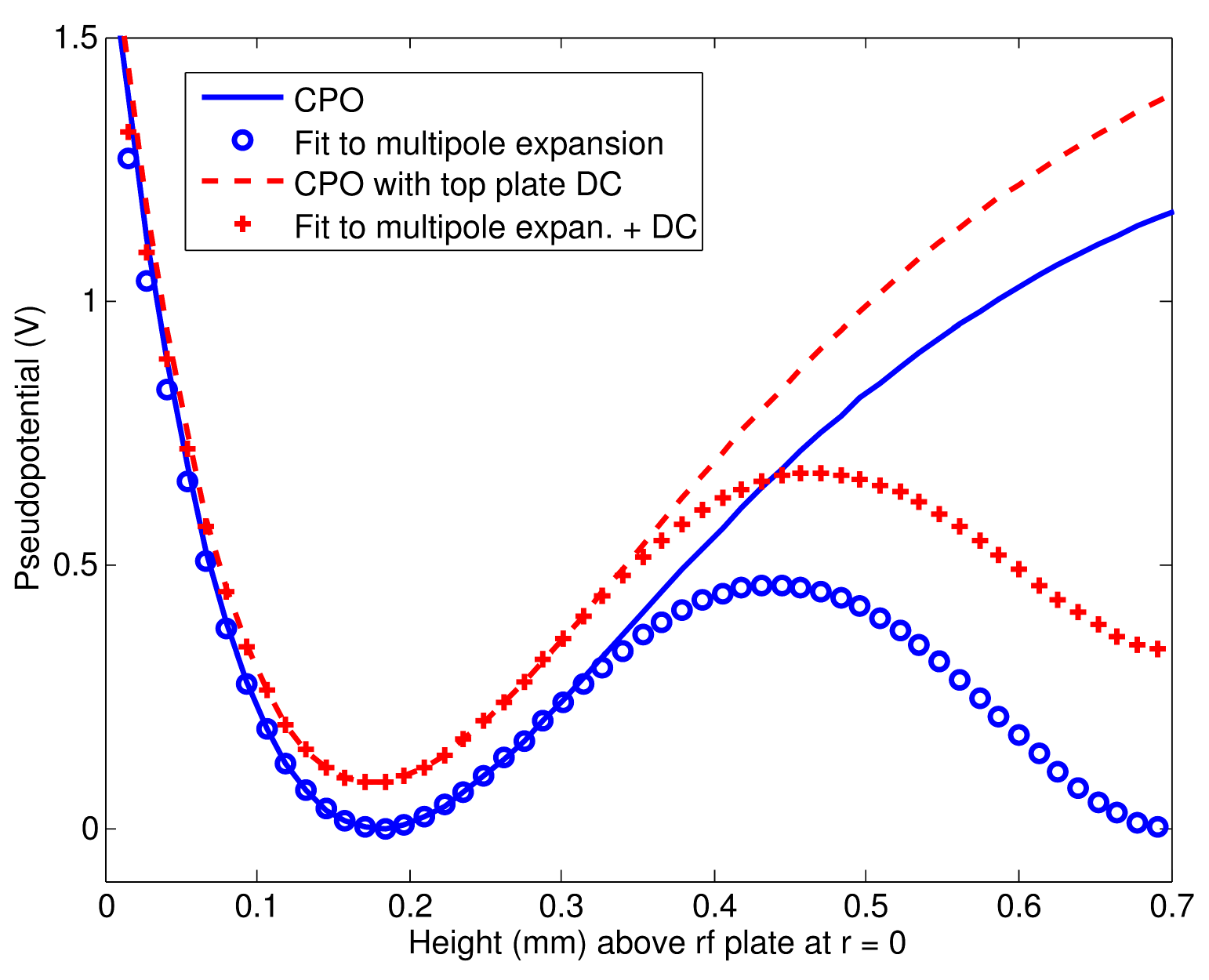}
\caption{Fit to Eq~\ref{eq:pseudopotential} of the CPO computed pseudopotential in the $z$ direction at the center of one well. The fit yields $r_1 = 3.1 \pm 0.1$ mm, $\alpha = -4.0 \pm 1.3$, and $z_1 = 19$~mm.}
\label{fig:pseudopotfit}
\end{center}
\end{figure}

The pseudopotential is given by
\begin{align}
	\Psi (\vec{x}) =  \frac{Q V^2}{m \Omega^2 r_1^4}&  \text{\LARGE [}r^2
	\left( 1 + \frac{3 \alpha z}{r_1} \right)^2 + \nonumber \\
	&4 z^2 \left( 1 + \frac{3 \alpha z}{2r_1} - \frac{ 3 \alpha r^2}{4 z r_1}
	\right)^2 \text{\LARGE]}.
\label{eq:pseudopot}
\end{align} From the pseudopotential we define secular frequencies which characterize the curvature of the pseudopotential in the harmonic region:
\begin{equation}
	\omega_z = 2\sqrt{2} \frac{QV}{m\Omega r_1^2}, \ \ \
	\omega_r = \sqrt{2} \frac{QV}{m \Omega r_1^2}
\end{equation} where $\omega_r$ is the secular frequency in the plane of the trap and $\omega_z$ is the secular frequency perpendicular to the plane of the trap. Note that $\omega_z/\Omega \approx q$ so that $\omega_z/\Omega$ gives a direct measure of the stability of the confined ions. The micromotion approximation is best when $\omega_z \ll \Omega$. 

An additional grounded plate may be added above the ions to shield them from stray charges, and a static potential $U$ may be applied to it. This change can be modeled by adding an extra $U(z-z_0)/z_1$ in the pseudopotential, where $z_1$  is a geometric factor with dimensions of length that depends on the height of this plane above the rf electrode and is computed, in practice, using numerical modeling. A consequence of this additional static potential is that $\omega_z$ is different:
\begin{equation}
	\omega_z^2 = 8\left(\frac{QV}{m \Omega r_1^2}
	\right)^2 \left(1 - \frac{144 \alpha m \Omega^2 r_1^3 U}{64 Q V^2 z_1} \right).
\label{eq:omega}
\end{equation}

 To obtain the constants $r_1$, $\alpha$, and $z_1$, we use the Charged Particle Optics (CPO) numerical modeling software package to model the trapping potentials. The lattice trap used for our experiments has a hole diameter of $h$~=~1.14 mm and a spacing between the centers of the holes of $d$~=~1.64 mm. A square section of the rf electrode measuring 10 lattice sites on each side was used for this modeling; for larger sections than this, the effect of adding additional sites on the potentials near the center was negligible. From a simulation of the trap, we obtain the value of the geometric factors: $r_1 = 3.1 \pm 0.1$ mm, $\alpha = -4.0 \pm 1.3$, and $z_1 = 19$~mm for a top plate 15~mm above the rf electrode. Errors arise from the nonlinear least-squares fit used to obtain $r_1$ and $\alpha$ from the (discrete) simulated potential. In Fig.~\ref{fig:pseudopotfit}, we compare the numerical potential for the lattice trap to  the analytical potential from the multipole expansion, indicating that near the minimum of a given potential well the multipole expansion gives an accurate approximation to the simulated pseudopotential.


\section{Atomic ion experiment}

\begin{figure}[b]
\begin{center}
\includegraphics[width=9cm]{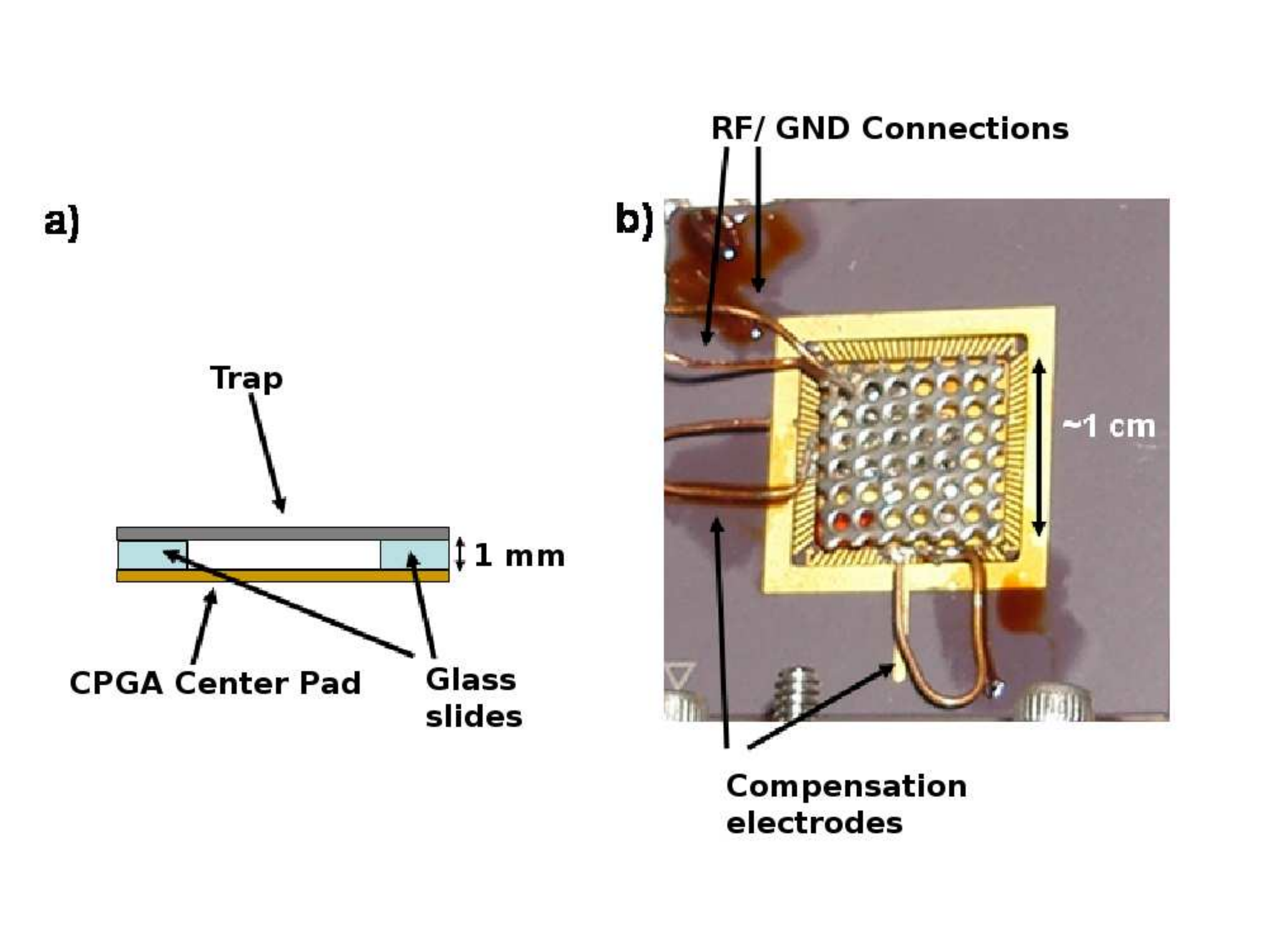}
\caption{(a) Schematic of the cross-section of the trap assembly. The trap electrode is held above the CPGA center pad on top of two 1~mm thick glass slides. (b) Photograph of the trap mounted in the CPGA. Connections for RF and GND are shown, as are the optional control electrodes for the $x$ and $y$ directions.}
\label{fig:atom_setup}
\end{center}
\end{figure}

We have observed stable confinement of $^{88}\mathrm{Sr}^+$ ions in a 6$\times$6 lattice trap, and verified the model of the trap discussed in Sec.~II by measuring the secular frequencies of the ions for one particular lattice site. The rf electrode is cut from a stainless steel mesh from Small Parts, Inc., Part No. PMX-045-A. It is mounted 1~mm above a grounded gold electrode on a ceramic pin grid array (CPGA) chip carrier (Fig.~\ref{fig:atom_setup}). An additional planar electrode (the ¨top plate¨) is mounted~1 cm above the rf electrode, to help shield the ions from stray charges. Electrical connections to both the rf and ground electrodes were made using a UHV-compatible solder from Accu-Glass (part no. 110796). The vacuum chamber was baked out to a base pressure of 7$\times 10^{-10}$ torr.

\begin{figure}[t]
\begin{center}
\includegraphics[width=9cm]{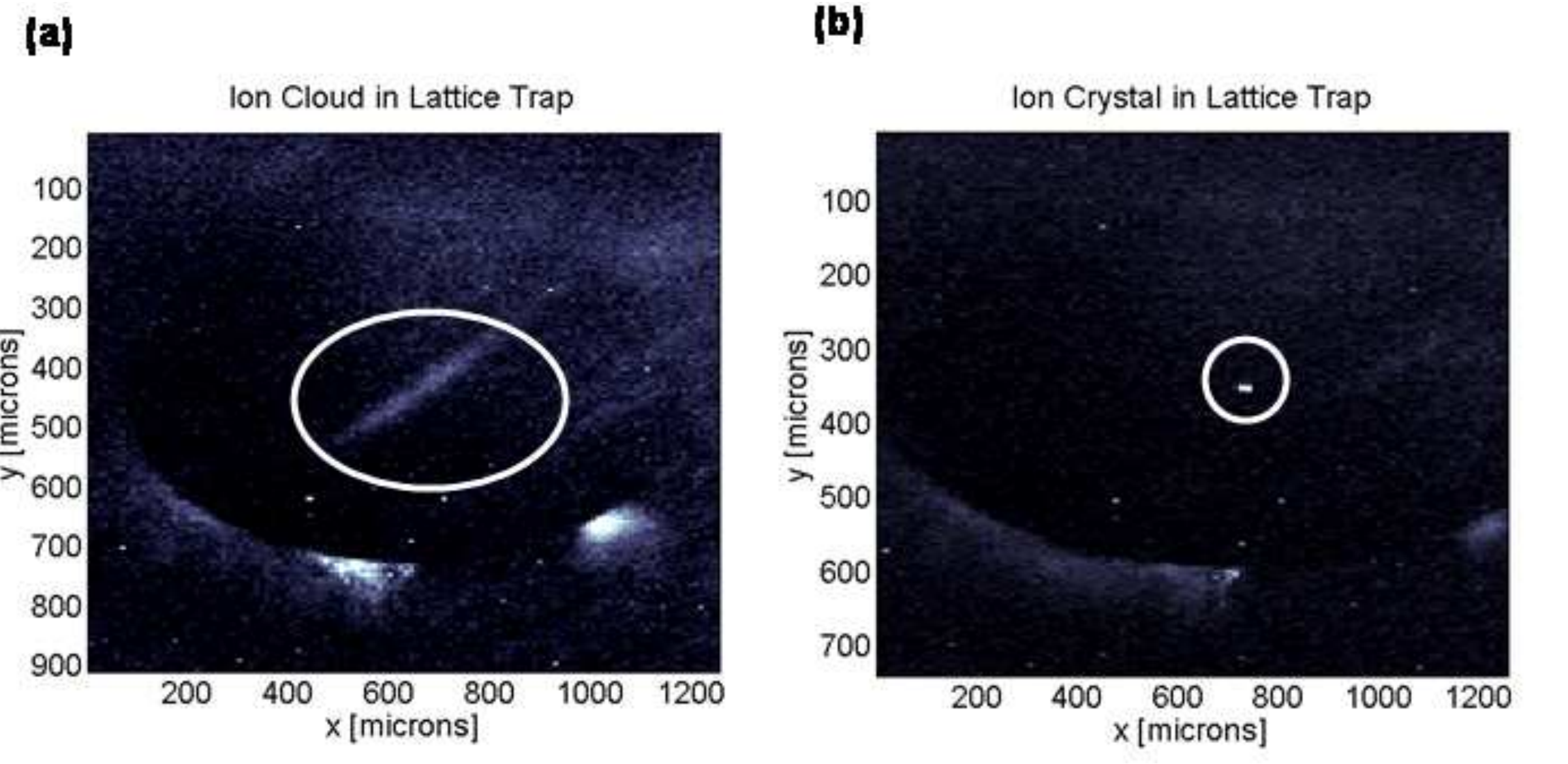}
\caption{(a) A cloud of ions (circled) intersects the detection lasers traversing the trapping region. The bright spots beneath the ions are laser scatter. (b) An ion crystal (circled) with a lifetime of $O$(15 minutes) is observed in the trap.}
\label{fig:atom_images}
\end{center}
\end{figure}

The trap is loaded with $^{88}\mathrm{Sr}^+$ by photoionizing a beam of neutral strontium produced by a resistive oven. This is a two-photon process at 460 nm and 405 nm that has been discussed in \cite{Brownutt:07}. Resonance fluorescence is imaged onto a CCD camera by simultaneously addressing the main 422 nm transition and the 1092 nm repumper. Typical laser powers used are 10 $\mu$W of 422 and 50 $\mu$W of 1092. Ions were observed as both clouds and crystals (Fig.~\ref{fig:atom_images}). The cloud lifetime is quite short ($O$(10 s)), but a small crystal has been kept in the trap, illuminated with cooling light, for up to 15 minutes. This short lifetime is attributable to the vacuum pressure.

A typical voltage of $V$~= 300~V at $\Omega/2\pi$~=~7.7~MHz was applied to the rf electrode using a power amplifier and helical resonator. Numerical modeling of the resulting pseudopotential yields secular frequency values of $\omega_r/2\pi$~= 300~kHz and $\omega_z/2\pi$~= 600~kHz. In order to test the model, we measure both secular frequencies as functions of the applied rf voltage $V$. We also compute a trap depth of 0.3~eV, which is the energy required for an ion at the potential minimum to escape.

\begin{figure}[h]
\begin{center}
\includegraphics[width=9cm]{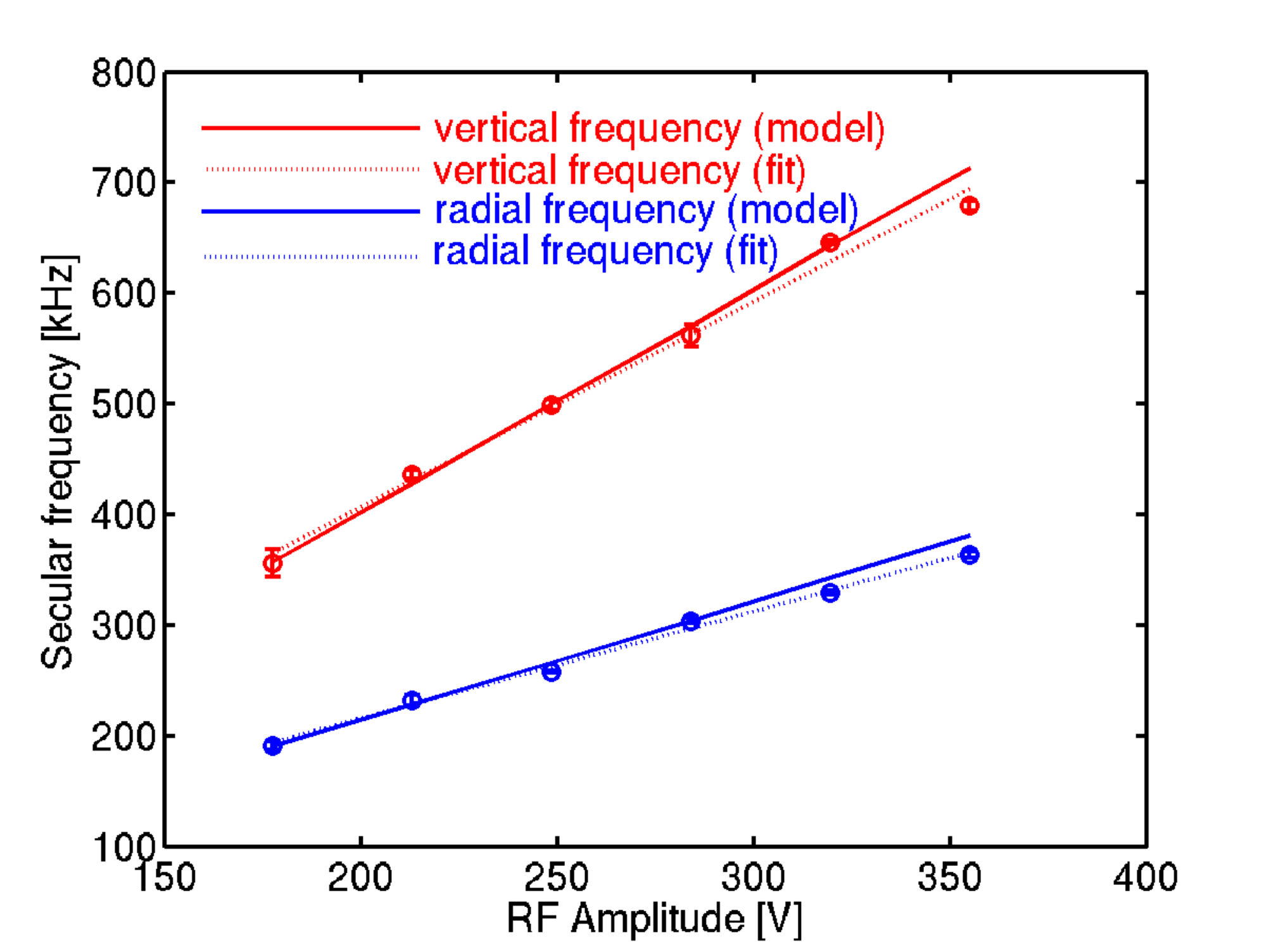}
\caption{Secular frequencies as a function of rf voltage for one site of the lattice trap. Circles represent data points, dotted lines represent linear fits to the data, and the solid lines are the predicted values from the model. The upper (red online) data are values of $\omega_z/2\pi$, and the lower (blue online) are values of $\omega_r/2\pi$.}
\label{fig:atom_secfrq}
\end{center}
\end{figure}

Secular frequencies were measured for one site near the center of the lattice using the standard method of applying a low-amplitude ($\sim$0.02~V) oscillating voltage to the top plate at the motional frequency of the ions. When each vibrational mode of the ions is stimulated, their heating causes measurable drops in the fluorescence intensity. This experiment was performed and compared to the model for several values of the drive voltage (Fig.~\ref{fig:atom_secfrq}). Agreement is very good; measured data points differ from the predicted values by at most 5\%, an error that results mainly from the approximation of the trap electrodes as perfect two-dimensional conductors for simulation. 

Although other sites near the center were also loaded, secular frequency measurements are presented here for only one site of the lattice. These experiments answer our questions regarding the ability to construct, operate, and accurately model a two-dimensional lattice ion trap: our agreement of less than 5\% could be made better by refining the simulation methods. 

\section{Macroion Experiment}

Another important test of the applicability of this lattice design to quantum simulations is the strength of interactions between ions in different wells. The charge to mass ratio of the strontium ions is unsuitable for this measurement in a lattice of this ($d$ = 1.64~mm) spacing. Instead, a lattice trap was loaded with ``macroions,'' aminopolystyrene microspheres with diameter $0.44$~$\mu$m (Spherotech Part No. AP-05-10).  The charge-to-mass ratio $Q/m$ of macroions used in the experiment leads to observable repulsions between ions in neighboring wells, although it takes on a relatively wide range of values due to the fact that $Q/m$ is not the same for every macroion. The use of macroions is also experimentally much less demanding than atomic ion trapping, because UHV pressures and laser cooling are not required. In fact, ions can be trapped in atmospheric pressure more easily than under vacuum, since air damping of ion motion increases the range of parameters suitable for stable trapping  \cite{Pearson:06,Winter:91,PearsonThesis:06}. 

\begin{figure}
\begin{center}
\begin{minipage}{.55\textwidth}
$\begin{array}{@{\hspace{-.45in}}c@{\hspace{.15in}}c}
	\mbox{(a)}
	\includegraphics[width=1.85in]{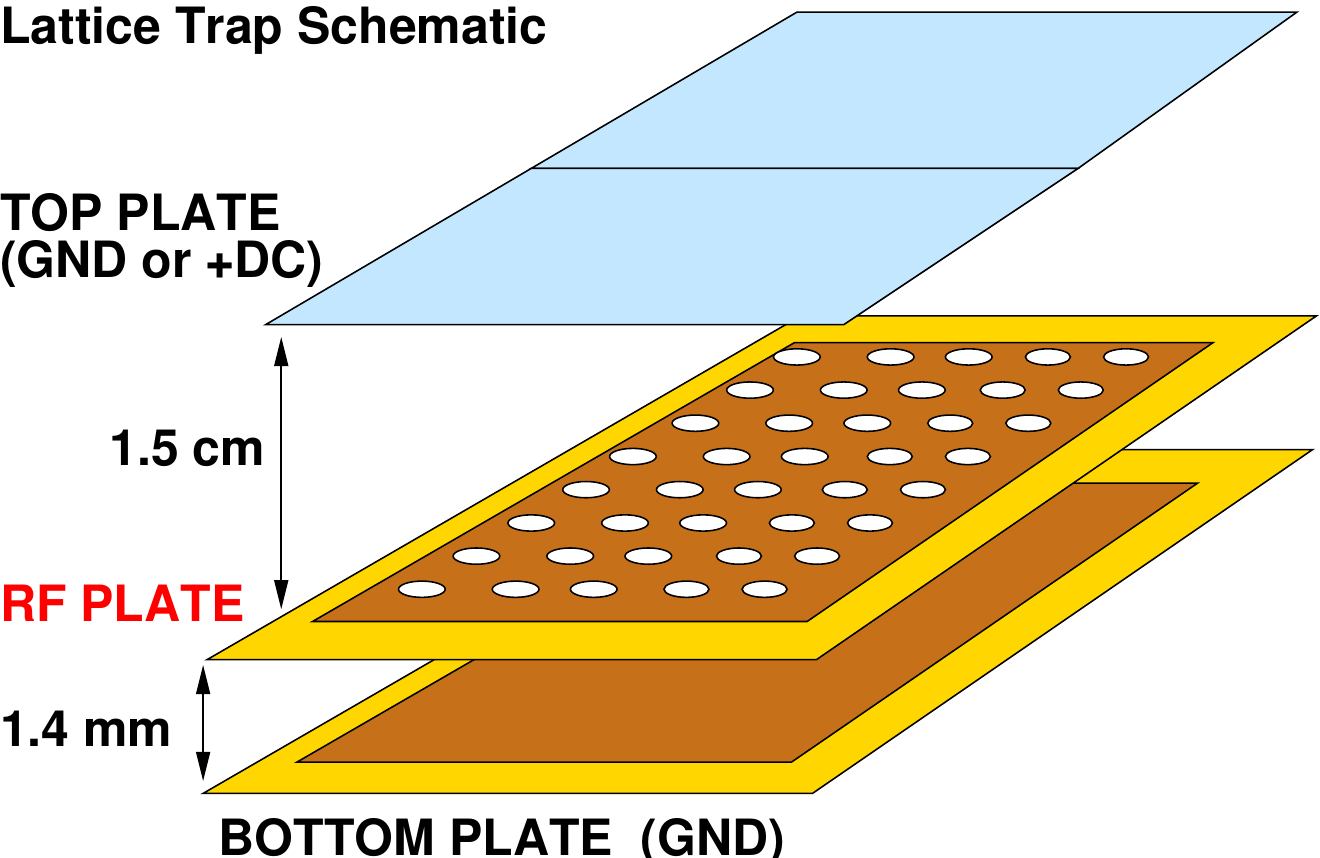}
	\mbox{ (b)}
	\includegraphics[width=1.4in]{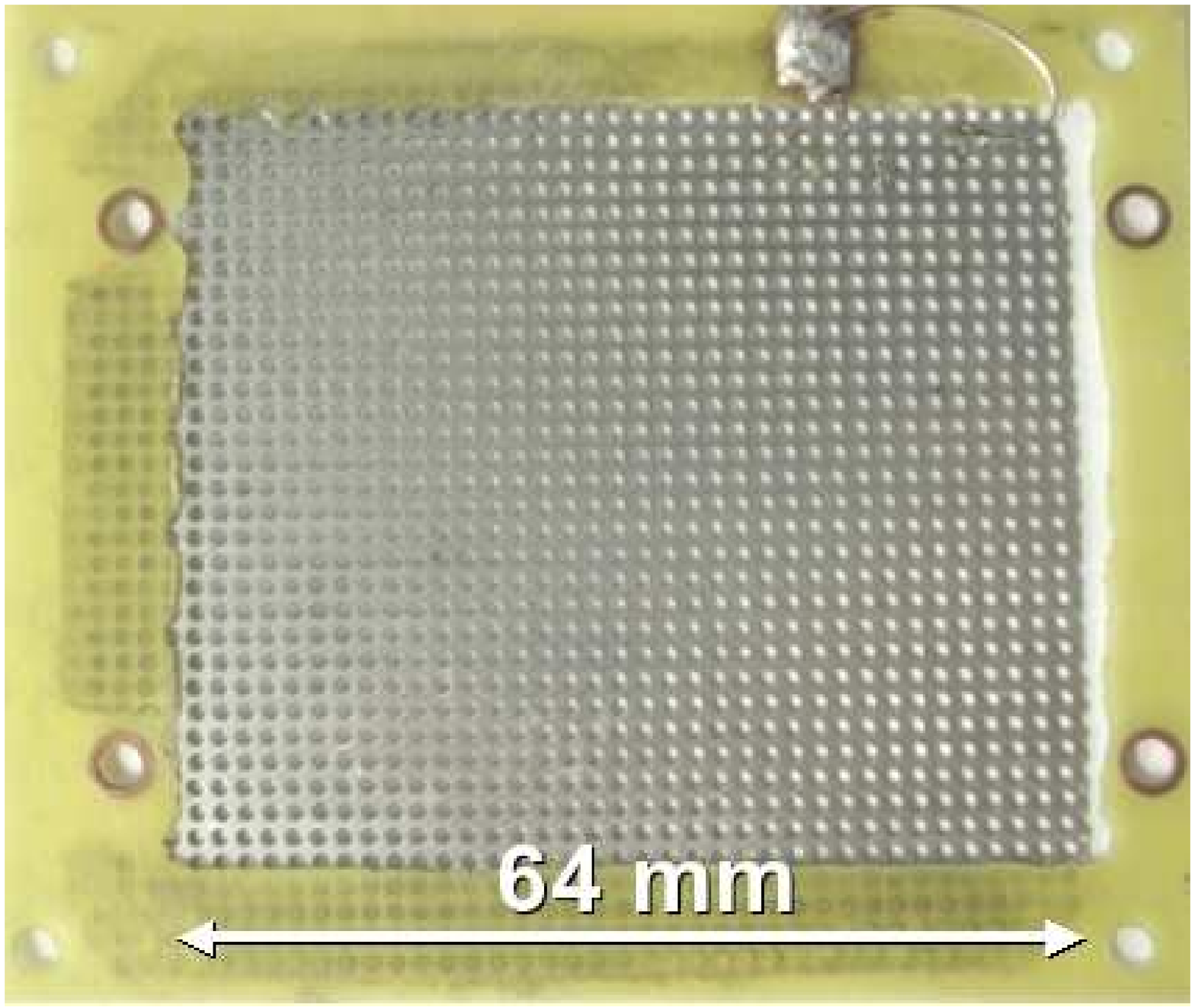}
\end{array}$
\end{minipage}
\end{center}
\caption{Experimental setup for the macroion experiment (Section IV). (a) 3-D schematic of lattice trap setup. (b) The lattice RF plate, as mounted for the microsphere experiment. The hole diameter is 1.14~mm and the hole spacing is 1.67~mm. The trap is supported by a printed circuit board.}
\label{fig:macroion_schematics}
\end{figure}

\begin{figure}[t]
\begin{center}
\includegraphics[width=.45\textwidth]{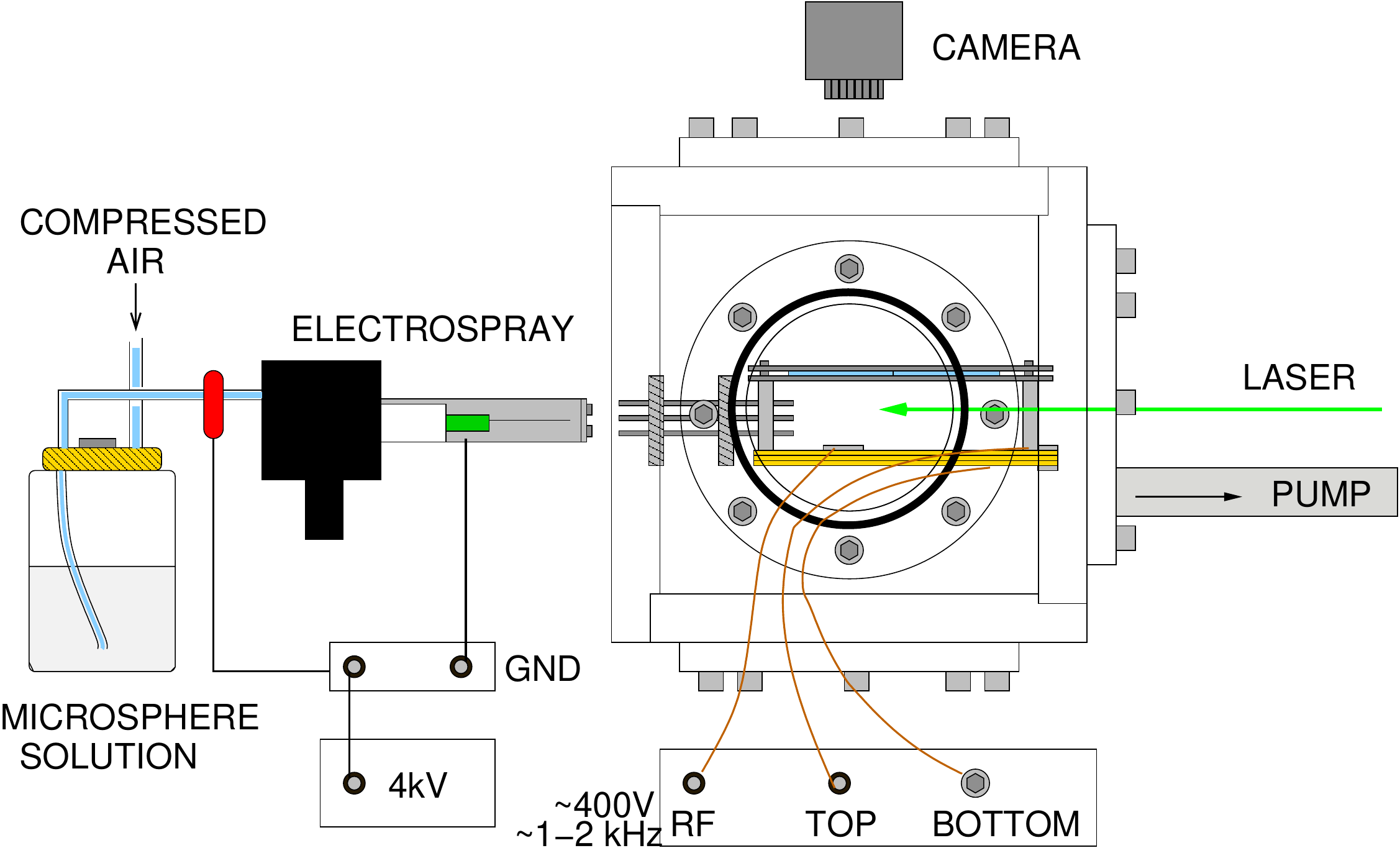}
\caption{Trapping apparatus. The lattice trap is inside a plastic chamber
which can be pumped down to $\approx$ 1 torr. Macroions are loaded via the electrospray and the 4-rod trap, which extends through one side of the chamber over the surface of the lattice trap.}
\label{fig:apparatus}
\end{center}
\end{figure}

\begin{figure}[t]
\begin{center}
\includegraphics[width=.45\textwidth]{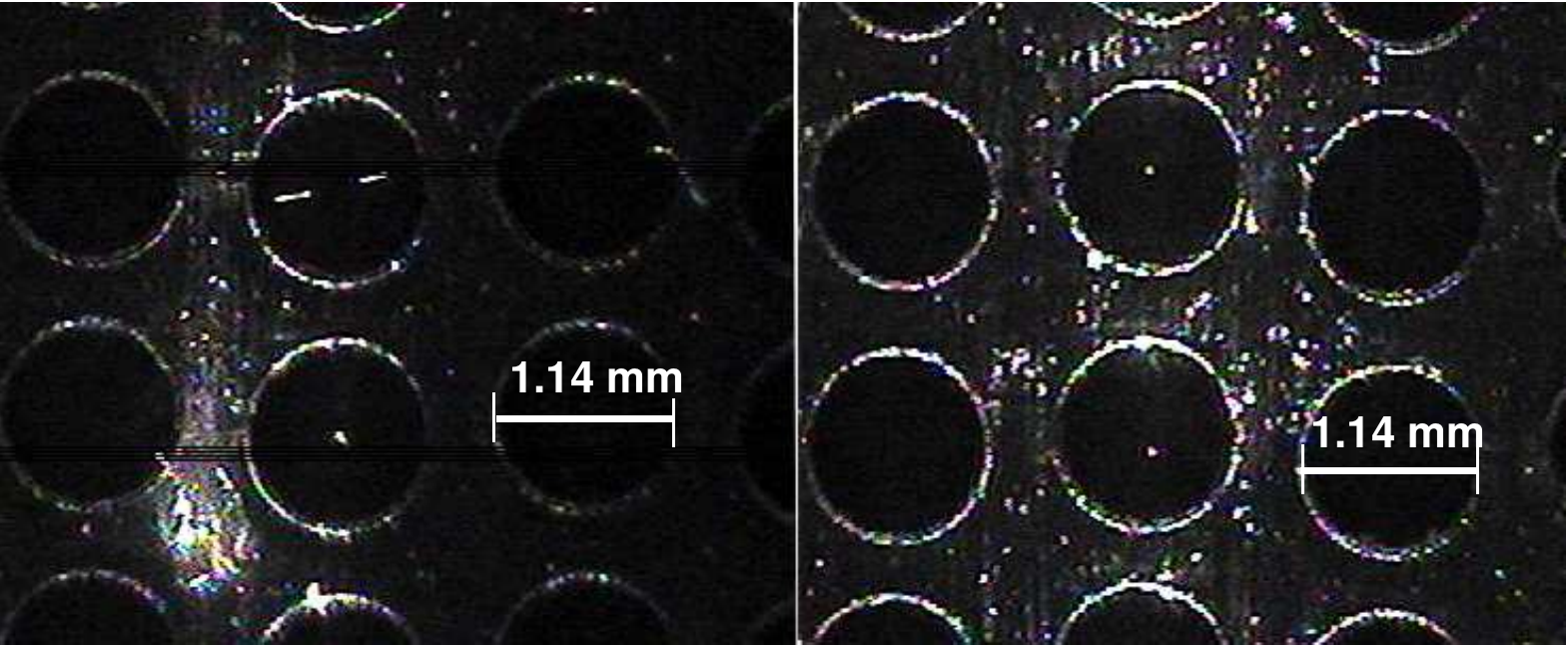}
\caption{Image from above of ions in the lattice. The dark holes are the holes in the rf electrode; the grounded plane is 1.4~mm beneath them. Single macroions appear as white dots that are levitated above the plane of the rf electrode and are illuminated by 532~nm  laser radiation at 5~mW. White dots on the surface of the rf electrode are due to stray light scatter. The left image was taken at $V$ = 300~V and $\Omega/2\pi$ = 1200~Hz and the right image was taken at $V$ = 300~V and $\Omega/2\pi$ = 1960~Hz. In the left figure, in the top well, two ions are shown repelling each other in the same well.}
\label{fig:topviewimage}
\end{center}
\end{figure}

Fig.~\ref{fig:apparatus} is a diagram of the experiment, which is an adaptation of the experiment in \cite{Pearson:06}. The main components of our apparatus are the electrospray system and the 4-rod loading trap. To load the ions in the lattice trap, we perform a modifcation of the method in \cite{Cai}, skipping the washing step. We prepared a buffer solution of 5~mL pure acetic acid, 26~mL 1M NaOH, and 5$\%$ suspension microsphere solution. The buffer solution reduces spread in macroion charge. We sonicated the solution for 10 minutes to mix the microspheres evenly in the solution, added 30 mL of methanol, and again sonicated for 10 minutes.

Compressed air, at a pressure of between 3 and 5~Psi, forces the buffer solution first through a 0.45~$\mu$m filter and then to the electrospray system. Here a copper wire at a voltage of 4~kV is inserted in the tubing and ionizes the solution as it passes. The ionized solution travels through a thin electrospray tip directed at a perforated, grounded electrospray plate and a 4-rod Paul trap just behind the electrospray plate. The electrospray tips were made from capillary tubes, which are heated and stretched to produce narrow openings of 75-125 $\mu$m.

As the solution enters the 4-rod trap, the methanol evaporates and the charged microspheres break into small clusters, the macroions. The 4-rod trap is driven at the drive parameters of the lattice trap and extends through the wall of a plastic chamber over the lattice trap. Inside the chamber, the 4-rod trap extends 0.75 cm over the lattice trap and the bottom rod of the 4-rod trap rests 1~mm above the ground plate.

The lattice trap is supported by standings inside the chamber, which can be closed on all sides to block air currents and can also be sealed and pumped down to $\sim$1 torr. Glass slides, which are coated with InTiO$_2$ so that one side is conductive, act as the top plate. They allow a top view of the trap, and are supported approximately 15~mm above the RF plate. The ions are then confined approximately 0.25 mm above the plane of the RF plate. An image of the ions in the trap is given in Fig.~\ref{fig:topviewimage}.

\begin{figure}[t]
\begin{center}
\begin{minipage}{0.4\textwidth}
\includegraphics[width=\textwidth]{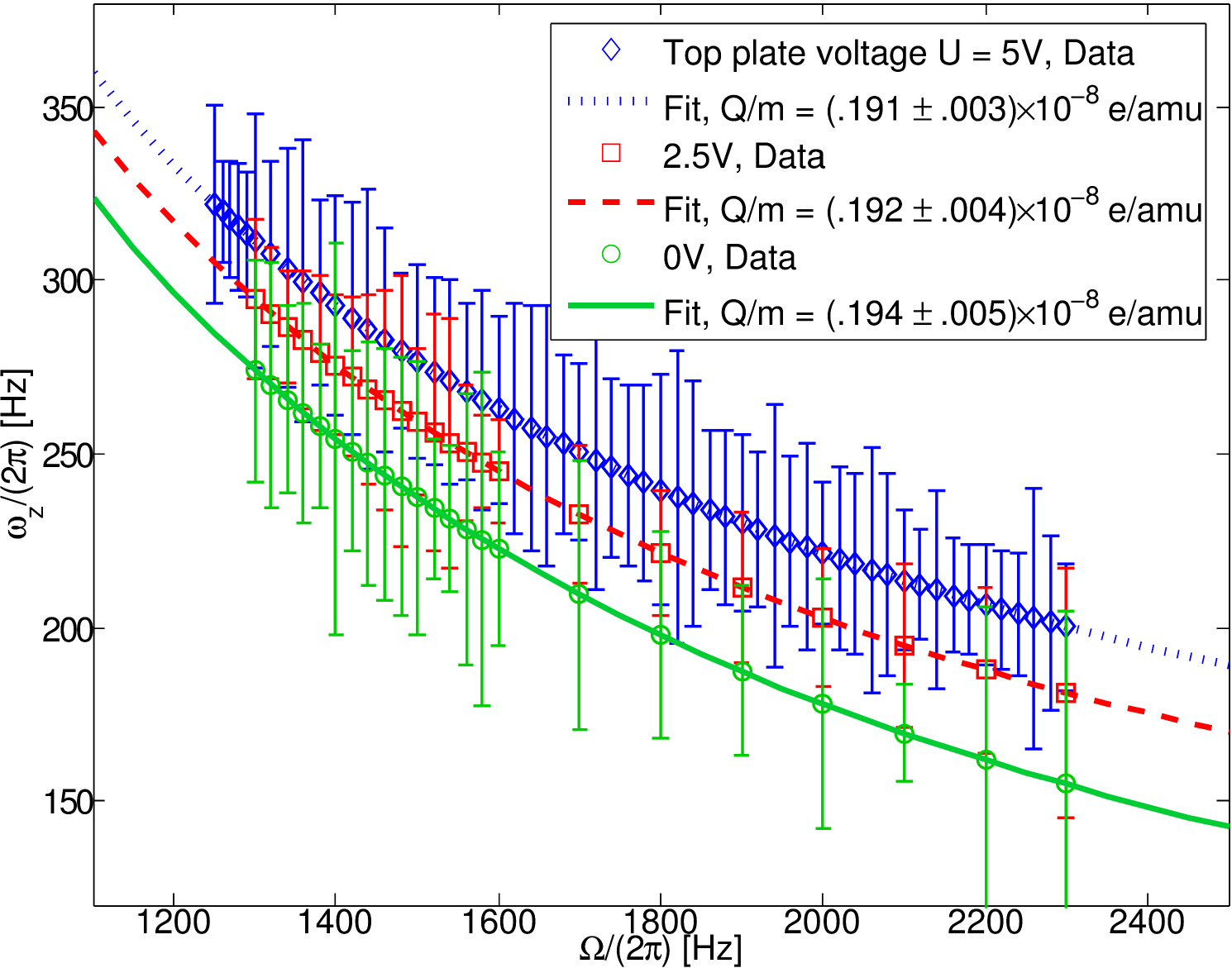}
\end{minipage}
\caption{$\omega_z$ vs. $\Omega$ for an isolated macroion at a drive voltage of $V$ = 255~V. The data for 0~V and 2.5~V come from the same ion. 
\\}
\label{fig:ionsepvswidth}
\end{center}
\end{figure}

Typical initial loading parameters for macroions were $\Omega/2\pi$ = 1000~Hz and $V$ = 250~V. We also applied a DC voltage of $U = 0-10$~V to the top plate to improve the trap depth. Before studying ion-ion repulsion, we estimate the $Q/m$ of the macroions by measuring their secular frequencies ($\omega_z$). To do this, we apply a low-amplitude tickle to the top plate and observe the resonances directly on a video camera as ions rapidly oscillate back and forth. A measurement of $\omega_z$ vs. $\Omega$ is shown in Fig.~\ref{fig:ionsepvswidth}. Using Eq.~\ref{eq:omega}, we fit these data to obtain a charge-to-mass ratio of $1.9 \times 10^{-9} e/$amu.



\begin{figure}
\begin{center}
\begin{minipage}{.4\textwidth}
\includegraphics*[width=\textwidth]{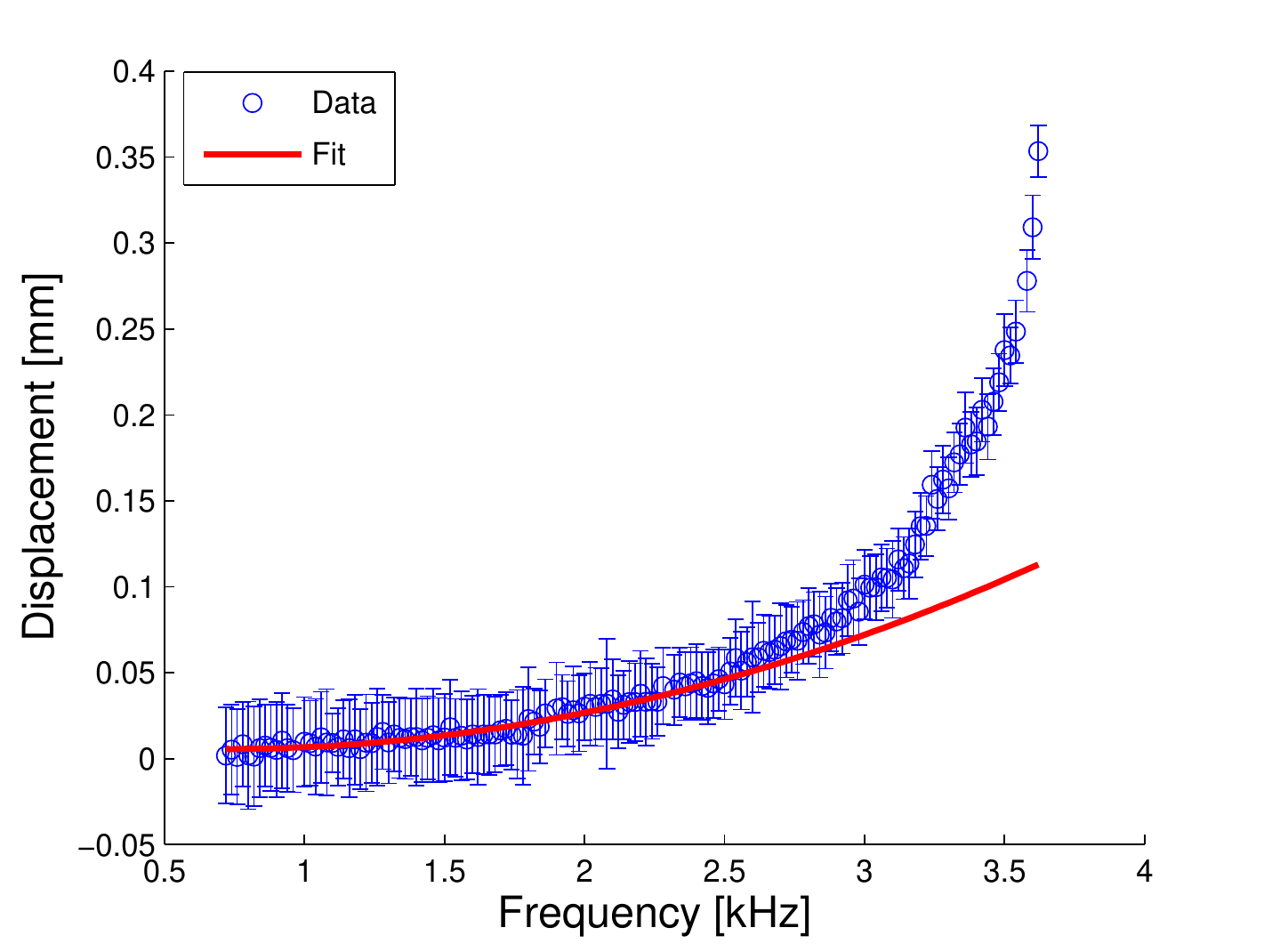}
\end{minipage}
\begin{minipage}{.4\textwidth}
\includegraphics*[width=\textwidth]{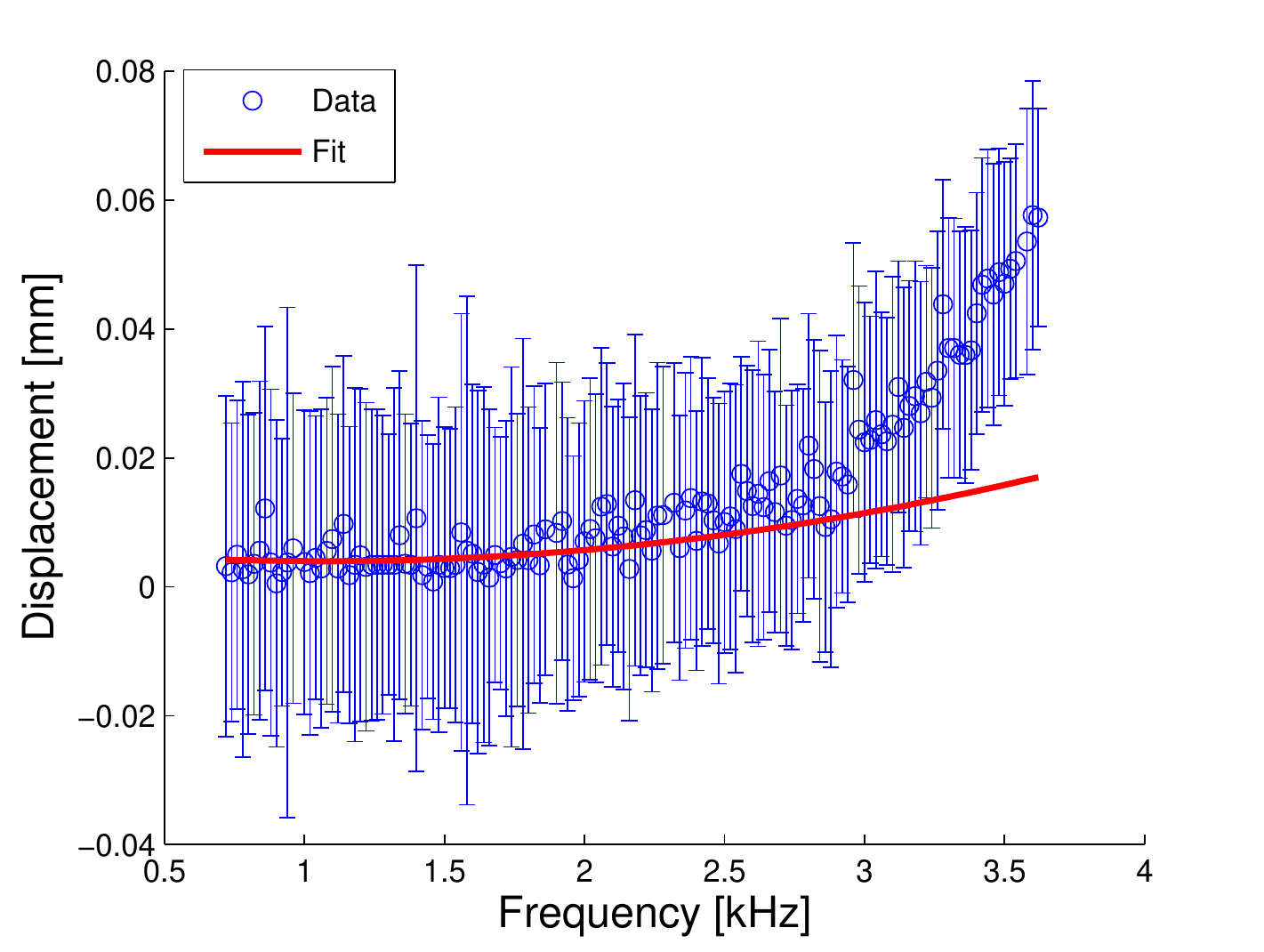}
\end{minipage}
\caption{\label{fig:sepwellfit_allpts} Ion displacement from the well center for two ions in neighboring wells, as a function of drive frequency. Due to charge asymmetry, the maximum displacements of the two ions differ by a factor of ten. The drive voltage is 350~V. The model breaks down for large displacements (high trap frequencies); fits only include data below $\Omega/2\pi$ = 2500~Hz.}
\label{fig:repulsion}
\end{center}
\end{figure}

We measured the Coulomb interaction of ions in neighboring lattice sites for six pairs of ions. In each pair, we measured the offset of each ion from the center of the well, as shown in Fig.~\ref{fig:topviewimage}. Note that while taking data on separation of two ions, we ascertained that wells adjacent to those containing the ions were all empty. The effect of a third ion in an adjacent well is significant.

A simple model of the interaction of two ions across wells is given as follows. An ion is confined by a force $-m \omega_r^2 x_1$, where $x_1$ is the ion offset from the center of the well.  Since generally $x_1 \ll d$, where $d$ is the lattice spacing, the ion is approximately repelled by a force $Q_1 Q_2/4 \pi \epsilon_0 sd^2$. Here $s$ is a screening factor and $Q_{1,2}$ are the charges of the first and second ion, respectively. The screening factor $s< 3$, where $s$~=3 for an ion sitting at a height .25~mm above an infinite conducting plane.

Using the expression for $\omega_r$ derived from Eq.~\ref{eq:pseudopot},
\begin{equation}
	x_1sd^2 \approx \Omega^2 \frac{ m r_1^4 Q_2}{8\pi \epsilon_0 V^2 Q_1}.
\label{eq:sepwellsmodel}
\end{equation} When $Q_1$ is not equal to $Q_2$, then the confining forces are characterized by different $\omega_r$ and the two ions have different offsets from equilibrium. The ratio in offsets, if the masses are comparable ($m_1 \approx m_2$), should be $(Q_2/Q_1)^2$. We observed exactly such an asymmetry between the offsets of the two ions, where typically $Q_2/Q_1$ is between 1 and 5. There may be additional small asymmetries due to edge effects and the presence of the 4-rod trap as well as differences in charge. Fig.~\ref{fig:repulsion} shows the displacement of a pair of ions as the rf drive frequency is varied, for one experimental run. The spread in charge to mass ratios and accordingly unknown values of $Q$ and $m$ for each ion (as in Ref.~\cite{Pearson:06}) does not permit us to compare the observed repulsion to a theoretical model. Nevertheless, we have answered our third question: ion-ion interaction in a mm-scale lattice trap is observable by the mutual Coulomb repulsion of the ions, albeit for a different system than the atomic ions that would be used for quantum simulation.

\section{Scaling Behavior}

The lattice trap discussed in this paper provides a fairly straightforward method for realizing a two-dimensional array of trapped ions. The question remains: how useful could this system be for quantum simulation of two-dimensional spin models \cite{Porras:04}, and in order to do this, what ion-ion spacing will be required? The scheme of Porras and Cirac is based on a pushing laser that exerts a state-dependent force on trapped ions that are coupled by their Coulomb interaction. In the limit in which the Coulomb interaction is small compared to the trapping potential (which is the case for lattice traps), the coupling rate between the ions is given by 

\begin{equation}
\label{eq:Jcoupling}
J = \frac{e^2 F^2}{8 \pi \epsilon_0 m^2 d^3 \omega^4}, 
\end{equation} where $F$ is the magnitude of the state-dependent force, $e$ is the charge of an electron, $m$ is the mass of each ion, $d$ is again the ion-ion distance, and $\omega$ is the trap secular frequency. For the lattice trap, $\omega$ of Eq.~\ref{eq:Jcoupling} is $\omega_r$. $F$ is assumed to be due to a tightly-focused laser beam, and arises from a spatially-dependent AC Stark shift. For a 5~W beam of 532~nm radiation that is focused from 50~$\mu$m to 3.5~$\mu$m over a distance of $d$ = 50~$\mu$m, in traps operating at $\omega$~=~2$\pi \cdot$250~kHz, we calculate a $J$ coupling of 1~kHz, which should be observable if the dominant decoherence time is greater than 2$\pi$/$J$. Similar values can be obtained by using less powerful lasers closer to the atomic resonance; we use the 532~nm beam as an example only because of the readily-available solid-state lasers at this wavelength. The motional decoherence rate expected in microfabricated surface-electrode traps becomes negligible relative to the internal state decoherence time if the trap is cooled to 6~K; rates for the former have been measured at as low as 5~quanta/s \cite{Labaziewicz:08}. Internal state decoherence times depend on the specific ion being used and also on classical controls, but coherence times as long as 10~s have been reported \cite{Langer:05, Haeffner:05}. 

Unfortunately, the scaling properties of lattice traps do not favor such a low secular frequency at small ion-ion spacings. To maintain trap stability, the $q$ parameter ($ q  = \frac{2 Q V}{m r_0^2 \Omega^2}$) must be held constant near 0.3 as $d$ varies. However, the trap depth $D \propto qV$ cannot be allowed to decrease too much, since traps of depth below about 100~meV have proven to be difficult to load. We note also that $r_1$ scales roughly as $d$. Therefore, the voltage must remain as high as it is for large traps, and the drive frequency $\Omega$ and secular frequency $\omega$ must increase as $1/d$, since $\omega \propto q\Omega$. According to Eq.~\eqref{eq:Jcoupling}, the increased trap frequency erases the gain of placing the ions closer together. Indeed, it appears in this regime that $J$  actually increases linearly with $d$, a result noted in Ref.~\cite{Chiaverini:08}. 

Greatly increasing the trap size is not only impractical, but may render the width of the ground state wave function of each trapped ion comparable to the laser wavelength, leaving the system outside of the Lamb-Dicke confinement regime. While some gains might be made from using the stronger field gradients of a standing wave configuration for the `pushing" laser, it is clear that the scaling of $\omega$ with $1/d$ is a discouraging feature of lattice traps. Of course, other options should be explored; a pressing question is how to modify the lattice trap design to allow for low motional frequencies for the directions along which the ions interact. One simple possibility might be to decrease the drive voltage $V$ (and consequently the trap depth) once the trap is loaded with ions and they have been laser-cooled to a temperature much lower than the trap depth. 

\section{Conclusions}

In conclusion, we have proposed a design for a layered planar rf lattice ion trap which contains many of the features desirable for quantum simulation, including the ability to control the structure of the 2D lattice of ions. We mentioned above that in the present design the structure of the ion lattice can be controlled by the intersection of the Doppler cooling, photoionization, and atomic beams; however, future realizations of the trap might also include individually controllable dc electrodes beneath each lattice site that could be used to eject unwanted ions. Also, in a future trap, the rf electrode could be specially fabricated to include only desired lattice sites. 

Of course, the square lattice used in this paper is only one possible geometry. Other lattices, including hexagonal ones, could also be used for quantum simulation. This would enable the possibility of observing spin frustration. As a first experiment, we envision trapping a triangular array of three ions and generating a spin-frustrated ground state. Although we have chosen to focus on the Porras-Cirac type spin model simulations, many of the schemes proposed for performing 2-D quantum simulations could be realized in such a trap \cite{Jane:02, Milburn:05, Porras:04,Porras:04BEC, Deng:05, Porras:06}. 

Our implementation in a mm-scale lattice trap is the first demonstration of stable confinement of ions in such a trap, and meaurement of the secular frequencies has confirmed our theoretical models of the trap. These were essential questions to address before we can move forward towards traps of this type on the scale of tens of microns, at which point they could become useful for quantum simulation (provided the secular frequencies can be kept low enough, as discussed above). It is also crucial to be able to measure interactions between ions, which we have done using charged microspheres. We hope that this work will stimulate further efforts towards measuring interactions between atomic ions in a lattice ion trap, paving the way for two-dimensional quantum simulations. 

We gratefully acknowledge funding from the MIT Undergradute Research Opportunities program, as well as fruitful discussions and laboratory assistance from Waseem Bakr, Christopher Pearson, Grace Cheung, and Ziliang Lin. 

\bibliographystyle{apsrev}

\begin{thebibliography}{36}
\expandafter\ifx\csname natexlab\endcsname\relax\def\natexlab#1{#1}\fi
\expandafter\ifx\csname bibnamefont\endcsname\relax
  \def\bibnamefont#1{#1}\fi
\expandafter\ifx\csname bibfnamefont\endcsname\relax
  \def\bibfnamefont#1{#1}\fi
\expandafter\ifx\csname citenamefont\endcsname\relax
  \def\citenamefont#1{#1}\fi
\expandafter\ifx\csname url\endcsname\relax
  \def\url#1{\texttt{#1}}\fi
\expandafter\ifx\csname urlprefix\endcsname\relax\def\urlprefix{URL }\fi
\providecommand{\bibinfo}[2]{#2}
\providecommand{\eprint}[2][]{\url{#2}}

\bibitem[{\citenamefont{Sachdev}(1999)}]{Sachdev:book}
\bibinfo{author}{\bibfnamefont{S.}~\bibnamefont{Sachdev}},
  \emph{\bibinfo{title}{{Quantum Phase Transitions}}}
  (\bibinfo{publisher}{{Cambridge University Press}},
  \bibinfo{address}{{Cambridge}}, \bibinfo{year}{1999}).

\bibitem[{\citenamefont{Diep}(1994)}]{Diep:book}
\bibinfo{editor}{\bibfnamefont{H.~T.} \bibnamefont{Diep}}, ed.,
  \emph{\bibinfo{title}{{Magnetic Systems with Competing Interactions}}}
  (\bibinfo{publisher}{{World Scientific}}, \bibinfo{address}{{Singapore}},
  \bibinfo{year}{1994}).

\bibitem[{\citenamefont{Moessner and Sondhi}(2001)}]{Moessner:01}
\bibinfo{author}{\bibfnamefont{R.}~\bibnamefont{Moessner}} \bibnamefont{and}
  \bibinfo{author}{\bibfnamefont{S.~L.} \bibnamefont{Sondhi}},
  \bibinfo{journal}{Phys. Rev. B} \textbf{\bibinfo{volume}{63}},
  \bibinfo{pages}{224401} (\bibinfo{year}{2001}).

\bibitem[{\citenamefont{Jian and Emig}(2005)}]{Jiang:05}
\bibinfo{author}{\bibfnamefont{Y.}~\bibnamefont{Jian}} \bibnamefont{and}
  \bibinfo{author}{\bibfnamefont{T.}~\bibnamefont{Emig}},
  \bibinfo{journal}{Phys. Rev. Lett.} \textbf{\bibinfo{volume}{94}},
  \bibinfo{pages}{110604} (\bibinfo{year}{2005}).

\bibitem[{\citenamefont{Somaroo et~al.}(1999)\citenamefont{Somaroo, Tseng,
  Havel, Laflamme, and Cory}}]{Cory:99}
\bibinfo{author}{\bibfnamefont{S.}~\bibnamefont{Somaroo}},
  \bibinfo{author}{\bibfnamefont{C.~H.} \bibnamefont{Tseng}},
  \bibinfo{author}{\bibfnamefont{T.~F.} \bibnamefont{Havel}},
  \bibinfo{author}{\bibfnamefont{R.}~\bibnamefont{Laflamme}}, \bibnamefont{and}
  \bibinfo{author}{\bibfnamefont{D.~G.} \bibnamefont{Cory}},
  \bibinfo{journal}{Phys. Rev. Lett.} \textbf{\bibinfo{volume}{82}},
  \bibinfo{pages}{5381} (\bibinfo{year}{1999}).

\bibitem[{\citenamefont{Negrevergne et~al.}(2004)\citenamefont{Negrevergne,
  Somma, Ortiz, Knill, and Laflamme}}]{Laflamme:04}
\bibinfo{author}{\bibfnamefont{C.}~\bibnamefont{Negrevergne}},
  \bibinfo{author}{\bibfnamefont{R.}~\bibnamefont{Somma}},
  \bibinfo{author}{\bibfnamefont{G.}~\bibnamefont{Ortiz}},
  \bibinfo{author}{\bibfnamefont{E.}~\bibnamefont{Knill}}, \bibnamefont{and}
  \bibinfo{author}{\bibfnamefont{R.}~\bibnamefont{Laflamme}},
  \bibinfo{journal}{Phys. Rev. A} \textbf{\bibinfo{volume}{71}},
  \bibinfo{pages}{032344} (\bibinfo{year}{2004}).

\bibitem[{\citenamefont{Peng et~al.}(2004)\citenamefont{Peng, De, and
  Suter}}]{Suter:04}
\bibinfo{author}{\bibfnamefont{X.}~\bibnamefont{Peng}},
  \bibinfo{author}{\bibfnamefont{J.}~\bibnamefont{De}}, \bibnamefont{and}
  \bibinfo{author}{\bibfnamefont{D.}~\bibnamefont{Suter}},
  \bibinfo{journal}{Phys. Rev. A} \textbf{\bibinfo{volume}{71}},
  \bibinfo{pages}{012307} (\bibinfo{year}{2004}).

\bibitem[{\citenamefont{Brown et~al.}(2006)\citenamefont{Brown, Clark, and
  Chuang}}]{Brown:06}
\bibinfo{author}{\bibfnamefont{K.~R.} \bibnamefont{Brown}},
  \bibinfo{author}{\bibfnamefont{R.}~\bibnamefont{Clark}}, \bibnamefont{and}
  \bibinfo{author}{\bibfnamefont{I.~L.} \bibnamefont{Chuang}},
  \bibinfo{journal}{Phys. Rev. Lett.} \textbf{\bibinfo{volume}{97}},
  \bibinfo{pages}{050504} (\bibinfo{year}{2006}).

\bibitem[{\citenamefont{Wineland et~al.}(1998)\citenamefont{Wineland, Monroe,
  Itano, King, Leibfried, Myatt, and Wood}}]{Wineland:98b}
\bibinfo{author}{\bibfnamefont{D.~J.} \bibnamefont{Wineland}},
  \bibinfo{author}{\bibfnamefont{C.}~\bibnamefont{Monroe}},
  \bibinfo{author}{\bibfnamefont{W.~M.} \bibnamefont{Itano}},
  \bibinfo{author}{\bibfnamefont{B.~E.} \bibnamefont{King}},
  \bibinfo{author}{\bibfnamefont{D.}~\bibnamefont{Leibfried}},
  \bibinfo{author}{\bibfnamefont{C.}~\bibnamefont{Myatt}}, \bibnamefont{and}
  \bibinfo{author}{\bibfnamefont{C.}~\bibnamefont{Wood}},
  \bibinfo{journal}{Physica Scripta} \textbf{\bibinfo{volume}{T76}},
  \bibinfo{pages}{147} (\bibinfo{year}{1998}).

\bibitem[{\citenamefont{Leibfried et~al.}(2002)\citenamefont{Leibfried,
  DeMarco, Meyer, Rowe, Ben-Kish, Britton, Itano, Jelenkovic, Langer, Rosenband
  et~al.}}]{Leibfried:02}
\bibinfo{author}{\bibfnamefont{D.}~\bibnamefont{Leibfried}},
  \bibinfo{author}{\bibfnamefont{B.}~\bibnamefont{DeMarco}},
  \bibinfo{author}{\bibfnamefont{V.}~\bibnamefont{Meyer}},
  \bibinfo{author}{\bibfnamefont{M.}~\bibnamefont{Rowe}},
  \bibinfo{author}{\bibfnamefont{A.}~\bibnamefont{Ben-Kish}},
  \bibinfo{author}{\bibfnamefont{J.}~\bibnamefont{Britton}},
  \bibinfo{author}{\bibfnamefont{W.~M.} \bibnamefont{Itano}},
  \bibinfo{author}{\bibfnamefont{B.}~\bibnamefont{Jelenkovic}},
  \bibinfo{author}{\bibfnamefont{C.}~\bibnamefont{Langer}},
  \bibinfo{author}{\bibfnamefont{T.}~\bibnamefont{Rosenband}},
  \bibnamefont{et~al.}, \bibinfo{journal}{Phys. Rev. Lett.}
  \textbf{\bibinfo{volume}{89}}, \bibinfo{pages}{247901}
  (\bibinfo{year}{2002}).

\bibitem[{\citenamefont{Friedenauer et~al.}(2008)\citenamefont{Friedenauer,
  Schmitz, Gl\"{u}ckert, Porras, and Sch\"{a}tz}}]{Schaetz:08}
\bibinfo{author}{\bibfnamefont{A.}~\bibnamefont{Friedenauer}},
  \bibinfo{author}{\bibfnamefont{H.}~\bibnamefont{Schmitz}},
  \bibinfo{author}{\bibfnamefont{J.}~\bibnamefont{Gl\"{u}ckert}},
  \bibinfo{author}{\bibfnamefont{D.}~\bibnamefont{Porras}}, \bibnamefont{and}
  \bibinfo{author}{\bibfnamefont{T.}~\bibnamefont{Sch\"{a}tz}}
  (\bibinfo{year}{2008}), \bibinfo{note}{eprint arxiv:0802.4072}.

\bibitem[{\citenamefont{Cataliotti et~al.}(2001)\citenamefont{Cataliotti,
  Burger, Fort, Maddaloni, Minardi, Trombettoni, Smerzi, and
  Inguscio}}]{Cataliotti:01}
\bibinfo{author}{\bibfnamefont{F.~S.} \bibnamefont{Cataliotti}},
  \bibinfo{author}{\bibfnamefont{S.}~\bibnamefont{Burger}},
  \bibinfo{author}{\bibfnamefont{C.}~\bibnamefont{Fort}},
  \bibinfo{author}{\bibfnamefont{P.}~\bibnamefont{Maddaloni}},
  \bibinfo{author}{\bibfnamefont{F.}~\bibnamefont{Minardi}},
  \bibinfo{author}{\bibfnamefont{A.}~\bibnamefont{Trombettoni}},
  \bibinfo{author}{\bibfnamefont{A.}~\bibnamefont{Smerzi}}, \bibnamefont{and}
  \bibinfo{author}{\bibfnamefont{M.}~\bibnamefont{Inguscio}},
  \bibinfo{journal}{Science} \textbf{\bibinfo{volume}{293}},
  \bibinfo{pages}{843} (\bibinfo{year}{2001}).

\bibitem[{\citenamefont{Paredes et~al.}(2004)\citenamefont{Paredes, Widera,
  Murg, Mandel, F\"{o}lling, Cirac, Shlyapnikov, H\"{a}nsch, and
  Bloch}}]{Paredes:04}
\bibinfo{author}{\bibfnamefont{B.}~\bibnamefont{Paredes}},
  \bibinfo{author}{\bibfnamefont{A.}~\bibnamefont{Widera}},
  \bibinfo{author}{\bibfnamefont{V.}~\bibnamefont{Murg}},
  \bibinfo{author}{\bibfnamefont{O.}~\bibnamefont{Mandel}},
  \bibinfo{author}{\bibfnamefont{S.}~\bibnamefont{F\"{o}lling}},
  \bibinfo{author}{\bibfnamefont{I.}~\bibnamefont{Cirac}},
  \bibinfo{author}{\bibfnamefont{G.~V.} \bibnamefont{Shlyapnikov}},
  \bibinfo{author}{\bibfnamefont{T.~W.} \bibnamefont{H\"{a}nsch}},
  \bibnamefont{and} \bibinfo{author}{\bibfnamefont{I.}~\bibnamefont{Bloch}},
  \bibinfo{journal}{Nature} \textbf{\bibinfo{volume}{429}},
  \bibinfo{pages}{277} (\bibinfo{year}{2004}).

\bibitem[{\citenamefont{Kinoshita et~al.}(2004)\citenamefont{Kinoshita, Wegner,
  and Weiss}}]{Weiss:04}
\bibinfo{author}{\bibfnamefont{T.}~\bibnamefont{Kinoshita}},
  \bibinfo{author}{\bibfnamefont{T.}~\bibnamefont{Wegner}}, \bibnamefont{and}
  \bibinfo{author}{\bibfnamefont{D.~S.} \bibnamefont{Weiss}},
  \bibinfo{journal}{Science} \textbf{\bibinfo{volume}{305}},
  \bibinfo{pages}{1125} (\bibinfo{year}{2004}).

\bibitem[{\citenamefont{Greiner et~al.}(2005)\citenamefont{Greiner, Mandel,
  Esslinger, H\"ansch, and Bloch}}]{Greiner:02}
\bibinfo{author}{\bibfnamefont{M.}~\bibnamefont{Greiner}},
  \bibinfo{author}{\bibfnamefont{O.}~\bibnamefont{Mandel}},
  \bibinfo{author}{\bibfnamefont{T.}~\bibnamefont{Esslinger}},
  \bibinfo{author}{\bibfnamefont{T.~W.} \bibnamefont{H\"ansch}},
  \bibnamefont{and} \bibinfo{author}{\bibfnamefont{I.}~\bibnamefont{Bloch}},
  \bibinfo{journal}{Nature} \textbf{\bibinfo{volume}{415}}, \bibinfo{pages}{39}
  (\bibinfo{year}{2005}).

\bibitem[{\citenamefont{Hofstetter et~al.}(2002)\citenamefont{Hofstetter,
  Cirac, Zoller, Demler, and Lukin}}]{Hofstetter:02}
\bibinfo{author}{\bibfnamefont{W.}~\bibnamefont{Hofstetter}},
  \bibinfo{author}{\bibfnamefont{J.~I.} \bibnamefont{Cirac}},
  \bibinfo{author}{\bibfnamefont{P.}~\bibnamefont{Zoller}},
  \bibinfo{author}{\bibfnamefont{E.}~\bibnamefont{Demler}}, \bibnamefont{and}
  \bibinfo{author}{\bibfnamefont{M.~D.} \bibnamefont{Lukin}},
  \bibinfo{journal}{Phys. Rev. Lett.} \textbf{\bibinfo{volume}{89}},
  \bibinfo{pages}{220407} (\bibinfo{year}{2002}).

\bibitem[{\citenamefont{Duan et~al.}(2003)\citenamefont{Duan, Demler, and
  Lukin}}]{Duan:03}
\bibinfo{author}{\bibfnamefont{L.-M.} \bibnamefont{Duan}},
  \bibinfo{author}{\bibfnamefont{E.}~\bibnamefont{Demler}}, \bibnamefont{and}
  \bibinfo{author}{\bibfnamefont{M.~D.} \bibnamefont{Lukin}},
  \bibinfo{journal}{Phys. Rev. Lett.} \textbf{\bibinfo{volume}{91}},
  \bibinfo{pages}{090402} (\bibinfo{year}{2003}).

\bibitem[{\citenamefont{S$\o$rensen et~al.}(2005)\citenamefont{S$\o$rensen,
  Demler, and Lukin}}]{Sorensen:05}
\bibinfo{author}{\bibfnamefont{A.~S.} \bibnamefont{S$\o$rensen}},
  \bibinfo{author}{\bibfnamefont{E.}~\bibnamefont{Demler}}, \bibnamefont{and}
  \bibinfo{author}{\bibfnamefont{M.~D.} \bibnamefont{Lukin}},
  \bibinfo{journal}{Phys. Rev. Lett.} \textbf{\bibinfo{volume}{94}},
  \bibinfo{pages}{086803} (\bibinfo{year}{2005}).

\bibitem[{\citenamefont{Jan\'e et~al.}(2003)\citenamefont{Jan\'e, Vidal, D\"ur,
  Zoller, and Cirac}}]{Jane:02}
\bibinfo{author}{\bibfnamefont{E.}~\bibnamefont{Jan\'e}},
  \bibinfo{author}{\bibfnamefont{G.}~\bibnamefont{Vidal}},
  \bibinfo{author}{\bibfnamefont{W.}~\bibnamefont{D\"ur}},
  \bibinfo{author}{\bibfnamefont{P.}~\bibnamefont{Zoller}}, \bibnamefont{and}
  \bibinfo{author}{\bibfnamefont{J.~I.} \bibnamefont{Cirac}},
  \bibinfo{journal}{Quant. Inf. and Comp.} \textbf{\bibinfo{volume}{3}},
  \bibinfo{pages}{15} (\bibinfo{year}{2003}).

\bibitem[{\citenamefont{Barjaktarevic et~al.}(2005)\citenamefont{Barjaktarevic,
  Milburn, and McKenzie}}]{Milburn:05}
\bibinfo{author}{\bibfnamefont{J.}~\bibnamefont{Barjaktarevic}},
  \bibinfo{author}{\bibfnamefont{G.}~\bibnamefont{Milburn}}, \bibnamefont{and}
  \bibinfo{author}{\bibfnamefont{R.~H.} \bibnamefont{McKenzie}},
  \bibinfo{journal}{Phys. Rev. A} \textbf{\bibinfo{volume}{71}},
  \bibinfo{pages}{012335} (\bibinfo{year}{2005}).

\bibitem[{\citenamefont{Porras and Cirac}(2004{\natexlab{a}})}]{Porras:04}
\bibinfo{author}{\bibfnamefont{D.}~\bibnamefont{Porras}} \bibnamefont{and}
  \bibinfo{author}{\bibfnamefont{J.~I.} \bibnamefont{Cirac}},
  \bibinfo{journal}{Phys. Rev. Lett.} \textbf{\bibinfo{volume}{92}},
  \bibinfo{pages}{207901} (\bibinfo{year}{2004}{\natexlab{a}}).

\bibitem[{\citenamefont{Porras and Cirac}(2004{\natexlab{b}})}]{Porras:04BEC}
\bibinfo{author}{\bibfnamefont{D.}~\bibnamefont{Porras}} \bibnamefont{and}
  \bibinfo{author}{\bibfnamefont{J.~I.} \bibnamefont{Cirac}},
  \bibinfo{journal}{Phys. Rev. Lett.} \textbf{\bibinfo{volume}{93}},
  \bibinfo{pages}{263602} (\bibinfo{year}{2004}{\natexlab{b}}).

\bibitem[{\citenamefont{Deng et~al.}(2005)\citenamefont{Deng, Porras, and
  Cirac}}]{Deng:05}
\bibinfo{author}{\bibfnamefont{X.-L.} \bibnamefont{Deng}},
  \bibinfo{author}{\bibfnamefont{D.}~\bibnamefont{Porras}}, \bibnamefont{and}
  \bibinfo{author}{\bibfnamefont{J.~I.} \bibnamefont{Cirac}},
  \bibinfo{journal}{Phys. Rev. A} \textbf{\bibinfo{volume}{72}},
  \bibinfo{pages}{063407} (\bibinfo{year}{2005}).

\bibitem[{\citenamefont{Porras and Cirac}(2006)}]{Porras:06}
\bibinfo{author}{\bibfnamefont{D.}~\bibnamefont{Porras}} \bibnamefont{and}
  \bibinfo{author}{\bibfnamefont{J.~I.} \bibnamefont{Cirac}},
  \bibinfo{journal}{Phys. Rev. Lett.} \textbf{\bibinfo{volume}{96}},
  \bibinfo{pages}{250501} (\bibinfo{year}{2006}).

\bibitem[{\citenamefont{Itano et~al.}(1998)\citenamefont{Itano, Bollinger, Tan,
  Jelenkovic, Mitchell, and Wineland}}]{Itano:98}
\bibinfo{author}{\bibfnamefont{W.~M.} \bibnamefont{Itano}},
  \bibinfo{author}{\bibfnamefont{J.~J.} \bibnamefont{Bollinger}},
  \bibinfo{author}{\bibfnamefont{J.~N.} \bibnamefont{Tan}},
  \bibinfo{author}{\bibfnamefont{B.}~\bibnamefont{Jelenkovic}},
  \bibinfo{author}{\bibfnamefont{T.~B.} \bibnamefont{Mitchell}},
  \bibnamefont{and} \bibinfo{author}{\bibfnamefont{D.~J.}
  \bibnamefont{Wineland}}, \bibinfo{journal}{Science}
  \textbf{\bibinfo{volume}{279}}, \bibinfo{pages}{686} (\bibinfo{year}{1998}).

\bibitem[{\citenamefont{Stahl et~al.}(2005)\citenamefont{Stahl, Galve, Alonso,
  Djekic, Quint, Valenzuela, Verdu, Vogel, and Werth}}]{Stahl:05}
\bibinfo{author}{\bibfnamefont{S.}~\bibnamefont{Stahl}},
  \bibinfo{author}{\bibfnamefont{F.}~\bibnamefont{Galve}},
  \bibinfo{author}{\bibfnamefont{J.}~\bibnamefont{Alonso}},
  \bibinfo{author}{\bibfnamefont{S.}~\bibnamefont{Djekic}},
  \bibinfo{author}{\bibfnamefont{W.}~\bibnamefont{Quint}},
  \bibinfo{author}{\bibfnamefont{T.}~\bibnamefont{Valenzuela}},
  \bibinfo{author}{\bibfnamefont{J.}~\bibnamefont{Verdu}},
  \bibinfo{author}{\bibfnamefont{M.}~\bibnamefont{Vogel}}, \bibnamefont{and}
  \bibinfo{author}{\bibfnamefont{G.}~\bibnamefont{Werth}},
  \bibinfo{journal}{Eur. Phys. J. D} \textbf{\bibinfo{volume}{32}},
  \bibinfo{pages}{139} (\bibinfo{year}{2005}).

\bibitem[{\citenamefont{Chiaverini and W.~E.~Lybarger}(2008)}]{Chiaverini:08}
\bibinfo{author}{\bibfnamefont{J.}~\bibnamefont{Chiaverini}} \bibnamefont{and}
  \bibinfo{author}{\bibfnamefont{J.}~\bibnamefont{W.~E.~Lybarger}},
  \bibinfo{journal}{Phys. Rev. A} \textbf{\bibinfo{volume}{77}},
  \bibinfo{pages}{022324} (\bibinfo{year}{2008}).

\bibitem[{\citenamefont{{P.K. Ghosh}}(1995)}]{Ghosh:book}
\bibinfo{author}{\bibnamefont{{P.K. Ghosh}}}, \emph{\bibinfo{title}{{Ion
  Traps}}} (\bibinfo{publisher}{{Oxford Science Publications}},
  \bibinfo{year}{1995}).

\bibitem[{\citenamefont{Brownnutt et~al.}(2007)\citenamefont{Brownnutt,
  Letchumanan, Wilpers, Thompson, Gill, and Sinclair}}]{Brownutt:07}
\bibinfo{author}{\bibfnamefont{M.}~\bibnamefont{Brownnutt}},
  \bibinfo{author}{\bibfnamefont{V.}~\bibnamefont{Letchumanan}},
  \bibinfo{author}{\bibfnamefont{G.}~\bibnamefont{Wilpers}},
  \bibinfo{author}{\bibfnamefont{R.~C.} \bibnamefont{Thompson}},
  \bibinfo{author}{\bibfnamefont{P.}~\bibnamefont{Gill}}, \bibnamefont{and}
  \bibinfo{author}{\bibfnamefont{A.}~\bibnamefont{Sinclair}},
  \bibinfo{journal}{Appl. Phys. B} \textbf{\bibinfo{volume}{87}},
  \bibinfo{pages}{411} (\bibinfo{year}{2007}).

\bibitem[{\citenamefont{Pearson et~al.}(2006)\citenamefont{Pearson, Leibrandt,
  Bakr, Mallard, Brown, and Chuang}}]{Pearson:06}
\bibinfo{author}{\bibfnamefont{C.~E.} \bibnamefont{Pearson}},
  \bibinfo{author}{\bibfnamefont{D.~R.} \bibnamefont{Leibrandt}},
  \bibinfo{author}{\bibfnamefont{W.~S.} \bibnamefont{Bakr}},
  \bibinfo{author}{\bibfnamefont{W.~J.} \bibnamefont{Mallard}},
  \bibinfo{author}{\bibfnamefont{K.~R.} \bibnamefont{Brown}}, \bibnamefont{and}
  \bibinfo{author}{\bibfnamefont{I.~L.} \bibnamefont{Chuang}},
  \bibinfo{journal}{Phys. Rev. A} \textbf{\bibinfo{volume}{73}},
  \bibinfo{pages}{032307} (\bibinfo{year}{2006}).

\bibitem[{\citenamefont{Winter and Ortjohann}(1991)}]{Winter:91}
\bibinfo{author}{\bibfnamefont{H.}~\bibnamefont{Winter}} \bibnamefont{and}
  \bibinfo{author}{\bibfnamefont{H.~W.} \bibnamefont{Ortjohann}},
  \bibinfo{journal}{American Journal of Physics} \textbf{\bibinfo{volume}{59}},
  \bibinfo{pages}{807} (\bibinfo{year}{1991}).

\bibitem[{\citenamefont{Pearson}(2006)}]{PearsonThesis:06}
\bibinfo{author}{\bibfnamefont{C.~E.} \bibnamefont{Pearson}}, Master's thesis,
  \bibinfo{school}{Massachusetts Institute of Technology Department of Physics}
  (\bibinfo{year}{2006}).

\bibitem[{\citenamefont{Cai et~al.}(2002)\citenamefont{Cai, Peng, Kuo, Lee, and
  Chang}}]{Cai}
\bibinfo{author}{\bibfnamefont{Y.}~\bibnamefont{Cai}},
  \bibinfo{author}{\bibfnamefont{W.}~\bibnamefont{Peng}},
  \bibinfo{author}{\bibfnamefont{S.}~\bibnamefont{Kuo}},
  \bibinfo{author}{\bibfnamefont{Y.}~\bibnamefont{Lee}}, \bibnamefont{and}
  \bibinfo{author}{\bibfnamefont{H.}~\bibnamefont{Chang}},
  \bibinfo{journal}{Anal. Chem.} \textbf{\bibinfo{volume}{74}},
  \bibinfo{pages}{232} (\bibinfo{year}{2002}).

\bibitem[{\citenamefont{Labaziewicz et~al.}(2008)\citenamefont{Labaziewicz, Ge,
  Antohi, Leibrandt, Brown, and Chuang}}]{Labaziewicz:08}
\bibinfo{author}{\bibfnamefont{J.}~\bibnamefont{Labaziewicz}},
  \bibinfo{author}{\bibfnamefont{Y.}~\bibnamefont{Ge}},
  \bibinfo{author}{\bibfnamefont{P.}~\bibnamefont{Antohi}},
  \bibinfo{author}{\bibfnamefont{D.}~\bibnamefont{Leibrandt}},
  \bibinfo{author}{\bibfnamefont{K.}~\bibnamefont{Brown}}, \bibnamefont{and}
  \bibinfo{author}{\bibfnamefont{I.}~\bibnamefont{Chuang}},
  \bibinfo{journal}{Phys. Rev. Lett.} \textbf{\bibinfo{volume}{100}},
  \bibinfo{pages}{013001} (\bibinfo{year}{2008}).

\bibitem[{\citenamefont{Langer et~al.}(2005)\citenamefont{Langer, Ozeri, Jost,
  Chiaverini, DeMarco, Ben-Kish, Blakestad, Britton, Hume, Itano
  et~al.}}]{Langer:05}
\bibinfo{author}{\bibfnamefont{C.}~\bibnamefont{Langer}},
  \bibinfo{author}{\bibfnamefont{R.}~\bibnamefont{Ozeri}},
  \bibinfo{author}{\bibfnamefont{J.~D.} \bibnamefont{Jost}},
  \bibinfo{author}{\bibfnamefont{J.}~\bibnamefont{Chiaverini}},
  \bibinfo{author}{\bibfnamefont{B.}~\bibnamefont{DeMarco}},
  \bibinfo{author}{\bibfnamefont{A.}~\bibnamefont{Ben-Kish}},
  \bibinfo{author}{\bibfnamefont{R.~B.} \bibnamefont{Blakestad}},
  \bibinfo{author}{\bibfnamefont{J.}~\bibnamefont{Britton}},
  \bibinfo{author}{\bibfnamefont{D.~B.} \bibnamefont{Hume}},
  \bibinfo{author}{\bibfnamefont{W.~M.} \bibnamefont{Itano}},
  \bibnamefont{et~al.}, \bibinfo{journal}{Phys. Rev. Lett.}
  \textbf{\bibinfo{volume}{95}}, \bibinfo{pages}{060502}
  (\bibinfo{year}{2005}).

\bibitem[{\citenamefont{H\"{a}ffner et~al.}(2005)\citenamefont{H\"{a}ffner,
  Schmidt-Kaler, H\"{a}nsel, Roos, K\"{o}rber, Chwalla, Riebe, Benhelm, Rapol,
  Becher et~al.}}]{Haeffner:05}
\bibinfo{author}{\bibfnamefont{H.}~\bibnamefont{H\"{a}ffner}},
  \bibinfo{author}{\bibfnamefont{F.}~\bibnamefont{Schmidt-Kaler}},
  \bibinfo{author}{\bibfnamefont{W.}~\bibnamefont{H\"{a}nsel}},
  \bibinfo{author}{\bibfnamefont{C.~F.} \bibnamefont{Roos}},
  \bibinfo{author}{\bibfnamefont{T.}~\bibnamefont{K\"{o}rber}},
  \bibinfo{author}{\bibfnamefont{M.}~\bibnamefont{Chwalla}},
  \bibinfo{author}{\bibfnamefont{M.}~\bibnamefont{Riebe}},
  \bibinfo{author}{\bibfnamefont{J.}~\bibnamefont{Benhelm}},
  \bibinfo{author}{\bibfnamefont{U.~D.} \bibnamefont{Rapol}},
  \bibinfo{author}{\bibfnamefont{C.}~\bibnamefont{Becher}},
  \bibnamefont{et~al.}, \bibinfo{journal}{Appl. Phys. B}
  \textbf{\bibinfo{volume}{81}}, \bibinfo{pages}{151} (\bibinfo{year}{2005}).

\end{thebibliography}

\end{document}